\documentclass{aastex}

\begin{document}

\title{The Physics of Protoplanetesimal Dust Agglomerates \\
II. Low Velocity Collision Properties}
\author{Doreen Langkowski}
\affil{Institut f\"ur Geophysik und extraterrestrische Physik,
Technische Universit\"at zu Braunschweig, Mendelssohnstr. 3,
D-38106 Braunschweig, Germany}

\author{Jens Teiser}
\affil{Institut f\"ur Planetologie, Westf\"alische
Wilhelmsuniversit\"at M\"unster, Wilhelm-Klemm-Str. 10, D-48149
M\"unster, Germany} \email{j.teiser@uni-muenster.de}

\author{J\"urgen Blum}
\affil{Institut f\"ur Geophysik und extraterrestrische Physik,
Technische Universit\"at zu Braunschweig, Mendelssohnstr. 3,
D-38106 Braunschweig, Germany} \email{j.blum@tu-bs.de}

\slugcomment{{\bf Accepted by the Astrophysical Journal}}

\begin{abstract}
For the investigation of collisions among protoplanetesimal dust
aggregates, we performed microgravity experiments in which the
impacts of high-porosity mm-sized dust aggregates into 2.5
cm-sized high-porosity dust aggregates can be studied. The dust
aggregates consisted either of monodisperse spherical, of
quasi-monodisperse irregular or of polydisperse irregular
micrometer-sized dust grains and were produced by random ballistic
deposition with porosities between 85\% and 93\%. Impact
velocities ranged from $\sim 0.1 ~\rm m~s^{-1}$ to $\sim 3 ~\rm
m~s^{-1}$ and impact angles were almost randomly distributed. In
addition to the smooth surfaces of the target aggregates formed in
our experiments, we ``molded'' target aggregates such that the
radii of the local surface curvatures corresponded to the
projectile radii, decreasing the targets porosities to $80-85$\%.
The experiments showed that impacts into the highest-porosity
targets almost always led to sticking, whereas for the less porous
dust aggregates, consisting of monodisperse spherical dust grains,
the collisions with intermediate velocities and high impact angles
resulted in the bouncing of the projectile with a mass transfer
from the target to the projectile aggregate. Sticking
probabilities for the impacts into the ``molded`` target
aggregates were considerably decreased. For the impacts into
smooth targets, we measured the depth of intrusion and the crater
volume and could derive some interesting dynamical properties
which can help to derive a collision model for protoplanetesimal
dust aggregates. Future models of the aggregate growth in
protoplanetary disks should take into account non-central impacts,
impact compression, the influence of the local radius of curvature
on the collisional outcome and the possible mass transfer between
target and projectile agglomerates in non-sticking collisions.
\end{abstract}


\section{Introduction}
Young stars are surrounded by accretion disks. It is now widely
accepted that the formation of planetesimals, the km-sized
precursors of the terrestrial planets and of the cores of the gas
planets, is initiated by the process of agglomeration of
protoplanetary dust in these disks. For the young Solar System,
the circumstellar disk is also known by the name of ''solar
nebula``. Due to the decreasing rate of accretion, the gas disk
cools down and fine dust particles condense. Typical initial grain
sizes, as evidenced by astronomical observations of circumstellar
accretion disks \citep{kesetal06,przetal03}, primitive meteorites
from our own Solar System \citep{ker93}, and recent measurements
from comets \citep{haretal05,keletal05,hoeetal06} are in the
(sub-)micrometer size range.

The velocities of these dust grains relative to the gas disk are
caused by Brownian motion, gravity-induced drift motion and gas
turbulence and result in mutual collisions among the grains
\citep{weicuz93} which can lead to the growth of dust
agglomerates. When the collision velocities are sufficiently low,
the dust particles show a hit-and-stick behavior
\citep{popetal00,bluwur00,domtie97} and the dust agglomerates are
bound by weak van der Waals forces \citep{heietal99}. As a
consequence, fractal dust agglomerates are formed whose fractal
dimensions $D_{\rm f}$, defined by the relation between
agglomerate mass $m$ and size $s$, $m \propto s^{D_{\rm f}}$,
range from $D_{\rm f} \approx 1.4 \ldots 1.5$
\citep{bluetal00,krablu04,pasdom06} for Brownian motion-dominated
growth to $D_{\rm f} \approx 1.8 \ldots 1.9$ for growth caused by
drift motions \citep{bluetal98} or gas turbulence
\citep{wurblu98}. Due to the increasing agglomerate mass and the
consequentially increasing collision velocities for protoplanetary
dust \citep{weicuz93}, collisions will eventually lead to the
compaction of the agglomerates \citep{domtie97,bluwur00} so that
the fractal dimensions will increase to $D_{\rm f} = 3$
\citep{weicuz93,blu04}. Such agglomerates are, however, not
compact but can be quite porous. Simulation experiments suggest
that the volume filling factors, given by
\begin{equation}\label{vff}
    \phi = 1 - p =\frac{\rho}{\rho_0}~,
\end{equation}
with $p$, $\rho$ and $\rho_0$ being the porosity, the mass density
of the aggregate and of the solid grain material, respectively,
should be as low as $\phi \le 0.20 \ldots 0.33$ even if the
collision velocities are as high as 50 $\rm m~s^{-1}$
(\citet{blusch04}; \citet{bluetal06}, hereafter referred to as
paper I). As long as the collision velocities are $\lesssim 1 \rm
m~s^{-1}$, the maximum volume filling factor for loose particle
ensembles is $\phi = 0.07 \ldots 0.15$ (\citet{blusch04}; paper
I). Very low collision velocities could lead to even lower volume
filling factors of $\phi \approx 0.01$ \citep{ormetal07}.

For fractal dust agglomerates, concurring theoretical
\citep{domtie97} and experimental \citep{bluwur00} results show
that above a threshold velocity of $\sim 1 \rm \, m~s^{-1}$, dust
agglomerates do no longer stick together but bounce off and
fragment upon collision. For non-fractal dust aggregates, impact
experiments by \citet{wuretal05a} and \citet{wuretal05b}, who used
cm-sized porous projectiles and decimeter-sized porous targets,
showed that for high-porosity targets ($\phi = 0.12 \ldots 0.26$)
no sticking occurs in the velocity range $16.5 \ldots 37.5 \, \rm
m~s^{-1}$ and a crater is formed on the target which leads to a
mass loss from the target. For compacted targets ($\phi = 0.34$),
a mass gain of the target of $\sim 50$ percent of the projectile
mass could be found for impact velocities $\ge 13 \, \rm
m~s^{-1}$.

Not much is known about the collision behavior of high-porosity
dust agglomerate in the velocity regime around $1 \, \rm
m~s^{-1}$. Earlier experiments by \citet{blumue93}, who used
mm-sized dust agglomerates with $\phi = 0.26$ and collision
velocities in the range $0.15 \ldots 3.8 \, \rm m~s^{-1}$, showed
that none of the collisions between (almost) equal-sized
agglomerates led to sticking, while at the highest collision
velocities fragmentation dominated. With our new technology for
the formation of high-porosity macroscopic dust agglomerates
(\citet{blusch04}; paper I), realistic macroscopic dust
agglomerates with volume filling factors in the range $\phi = 0.07
\ldots 0.15$ are now available with which the collision behavior
of protoplanetary dust agglomerates can be investigated.

Sect. \ref{sectobj} gives an overview of the scientific objectives
of this work, Sect. \ref{sectsetup} presents the experimental
technology and the properties of the used dust agglomerates, Sect.
\ref{sectexpres} gives the experimental results on sticking
efficiencies, mass transfer, coefficients of restitution, energy
loss, tensile strength, crater formation and impact fragmentation.
Sect. \ref{sectdisc} discusses the low-velocity impact dynamics of
high-porosity dust agglomerates, Sect. \ref{sectsummary} gives a
summary of our experimental findings, and in Sect.
\ref{sectconclusion}, we draw conclusions from our work for the
formation of larger objects in the protoplanetary nebula.

\section{\label{sectobj}Objectives}
This paper describes novel low-velocity impact experiments between
high-porosity dusty projectiles of diameters between $\sim 0.2$ mm
and $\sim 3$ mm and high-porosity dusty targets of 2.5 cm
diameter. From previous modelling and laboratory work it became
clear that above a certain agglomerate size the hitherto present
growth of fractal agglomerates is no longer feasible due to the
increasing collision energy with increasing agglomerate size
\citep{domtie97,bluwur00}. Thus, the agglomerates above a certain
size limit should be non-fractal but highly porous in structure
\citep{blu04,ormetal07}. With a new experimental method, we are
able to manufacture macroscopic, high-porosity dust agglomerates
(see Sect. \ref{sectsetup}) with which realistic collision
experiments for macroscopic protoplanetesimals are possible. The
experiments described below have the following main scientific
objectives:
\begin{enumerate}
    \item Under what physical conditions do high-porosity dust
    agglomerates stick?
    \item What are the possible outcomes of collisions between
    high-porosity dust agglomerates?
    \item What is the influence of projectile mass, velocity,
    impact angle, local radius of curvature, and
    porosity on the outcome of a collision?
    \item What are the sticking efficiencies in mutual collisions
    between protoplanetesimal, non-fractal dust agglomerates?
    \item What are the coefficients of restitution in non-sticking
    collisions?
\end{enumerate}

\section{\label{sectsetup}Experimental Setup, Samples and Procedures}
For the experimental simulations of dust-dust collisions between
mm-sized and cm-sized dust agglomerates, we used monolithic
dust-agglomerate samples of 2.5 cm diameter and $\sim 1$ cm height
as targets and smaller fragments of such samples as projectiles.
All samples were formed by random ballistic deposition (RBD; see
paper I and \citet{blusch04} for details of the RBD process).
Table \ref{table1} summarizes the physical properties of the
particles and the resulting RBD agglomerates. Three monomer
particle types were used: (1) monodisperse $\rm SiO_2$ spheres
with 1.5 $\rm \mu m$ diameter, (2) quasi-monodisperse, irregular
diamond particles with $\sim 1.5$ $\rm \mu m$ diameter, and (3)
polydisperse, irregular $\rm SiO_2$ grains with diameters in the
range 0.1 - 10 $\rm \mu m$. The motivation for the use of these
three samples was less in their cosmochemical relevance for
protoplanetary dust but lay in their span of morphologies and size
distributions and the question how these morphologies influence
the outcome of the collisions. As was shown by \citet{popetal00},
the impact of the material on the sticking behavior is smaller
than the influence of the individual particle morphology. Our
experiments span a wide range in morphological parameters so that
we consider them relevant for protoplanetary collision processes.
\placetable{table1}

It is interesting to note that the volume filling factor of the
agglomerates (see Table \ref{table1}), which describes the
fraction of volume filled with particles, is dependent on the
particle morphology and the width of the particle size
distribution (paper I), and varies by a factor of two between
monodisperse, spherical monomers ($\phi = 0.15$) and polydisperse,
irregular grains ($\phi = 0.07$). The broader the size
distribution, the fluffier the agglomerates are. For RBD
agglomerates consisting of spherical monodisperse particles, one
expects a volume filling factor of $\phi = 0.15$
\citep{vol59,watetal97} which is exactly found in our samples (see
paper I and \citet{blusch04}).

\begin{figure}[htp]
  \plotone{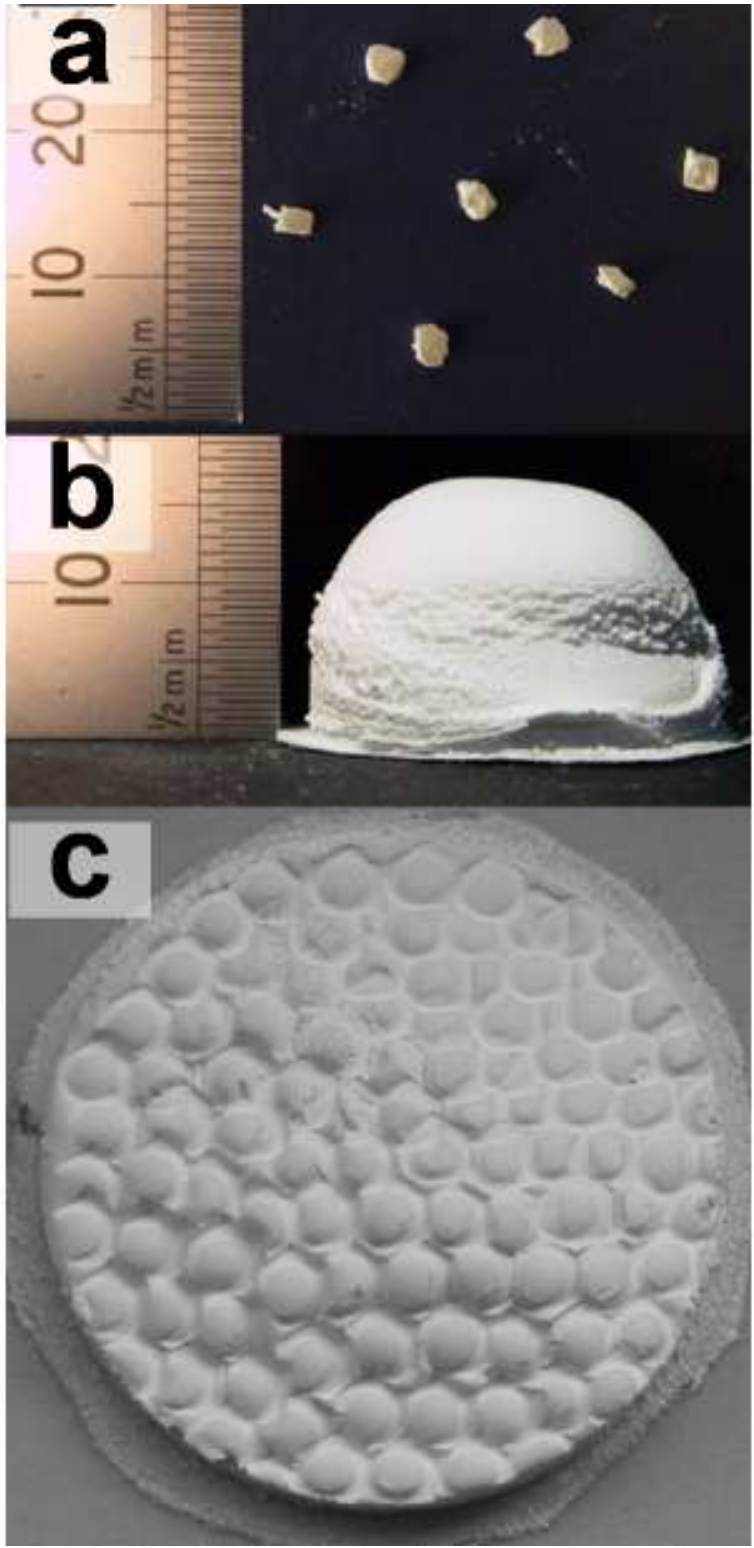}
  \figcaption{\label{targetprojectiles}(a) Example of mm-sized
  projectiles consisting of spherical $\rm SiO_2$ particles. (b)
  Example of an unprocessed target with 2.5 cm diameter
  consisting of spherical $\rm SiO_2$ particles. (c) Example of a
  ``molded'' target with 2.5 cm diameter
  consisting of spherical $\rm SiO_2$ particles.}
\end{figure}

To simulate the collisional history of protoplanetary dust
aggregates and to test the influence of the local target surface
curvature on the outcome of a collision, we also performed a
series of impact experiments into ``molded'' targets. These
targets, consisting of $1.5~\rm \mu m$ diameter $\rm SiO_2$
spheres with an initial volume filling factor of $\phi = 0.15$,
were slightly locally compacted by a half-spherical mold with 1 mm
radius so that they ultimately consisted of asperities with local
radii of curvature of 1 mm (representing, e.g., sticking
projectiles from previous impacts). X-ray tomography of the
``molded'' targets showed that the ``hills'' consisted of
uncompressed aggregated dust ($\phi = 0.15-0.17$), while the
``valleys'' were slightly compressed to a volume filling factor of
$\phi = 0.17-0.20$. Examples of an unprocessed and a processed
target and several projectile agglomerates are shown in Fig.
\ref{targetprojectiles}.

\begin{figure}[htp]
  \plotone{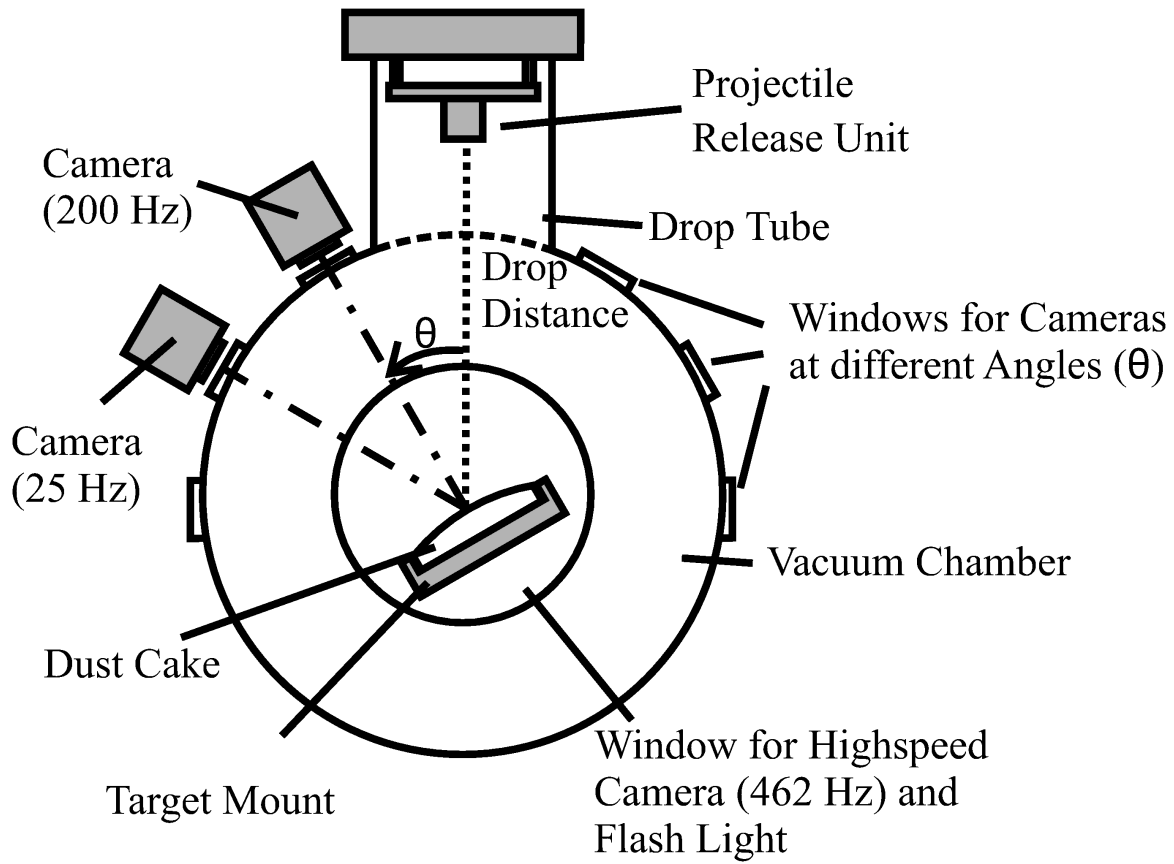}
  \figcaption{\label{setup}Schematics of the experimental
  setup used in this work.}
\end{figure}

For the realization of collisions between dusty targets and dusty
projectiles we developed a setup whose functionality is depicted
in Fig. \ref{setup}.

The experiments are performed under microgravity conditions within
a pressurized capsule which is, prior to each impact experiment,
held at the top of the microgravity drop tower in Bremen. Inside
the experimental setup, a set of five typically mm-sized
projectile agglomerates is held in a device which is able to
simultaneously release the projectiles when an electrical current
is applied to two solenoid magnets. The projectiles' release is
done while the experiment capsule is still held at ambient
gravitational acceleration. Due to the gravitational acceleration,
the released projectiles gain vertical velocities proportional to
their time of flight. After some preselected travel distance (or
time of flight), the experiment capsule is released to free fall.
Thus, the relative velocity between the projectiles and the target
is frozen and the impacts happen at a residual acceleration level
of less than $10^{-5} \rm ~m~s^{-2}$, mimicking the conditions in
protoplanetary disks. The free-fall height of the projectiles can
be adjusted from $\sim 5$ cm to $\sim 45$ cm. Due to a limited
acceleration length of the projectiles of $\lesssim 45$ cm, the
{\it maximum} impact velocity is $\sim 3 ~\rm m~s^{-1}$, while the
{\it actual} impact velocity is only determined by the time lapse
between projectile release and start of the free-fall phase of the
whole experiment. For the simulation of random impacts, the target
can be tilted by 30 and 60 degrees relative to the velocity vector
of the projectiles. For the exclusion of aerodynamic effects
during the collisions, the experiment chamber is evacuated to less
than 20 Pa pressure.

During the free-fall time of 4.74 seconds, the impacts are
observed by a set of different cameras and illumination schemes
(see Fig. \ref{setup}): (1) A high-speed (462 frames per second
[fps]), high-resolution (1k $\times$ 1k pixels) camera with a
field of view (FOV) of $30 \times 30 ~ \rm mm^2$ is arranged such
that it observes the impacts tangential to the target surface; the
illumination is provided by a synchronized Xe flash lamp with
$\sim 1 ~\rm \mu s$ flash duration. Due to the limited resolution,
this camera can only detect particles $> 30 \rm \mu m$. (2)
Another high-speed (220 fps), low-resolution (256 $\times$ 256
pixels) camera with a FOV of $56 \times 56 ~ \rm mm^2$ can observe
the impacts perpendicular (in the case of non-normal impacts) or
almost perpendicular (in the case of normal impacts) to the target
surface; additional illumination is provided by a laser curtain
with 30 mm width and 1 mm thickness parallel and close to the
surface of the target; this camera-illumination combination was
specifically installed for detecting small fragments that are
otherwise invisible. (3) An additional low-speed (25 fps),
low-resolution (720 $\times$ 576 pixels) video camera with a FOV
of $52 \times 41 ~ \rm mm^2$, located 30 degrees from the target
normal, observes the target surface prior and after the impacts; a
point-source illumination for the determination of, e.g., crater
depths is provided by a halogen lamp located 32 degrees above the
target's ``horizon''.

Data analysis consists of a thorough image analysis including the
determination of projectile sizes, impact velocities and impact
angles with respect to the local target normal, and the
determination of the outcomes of the collisions (sticking, rebound
or fragmentation) as well as parameters determining the effects
during the collisions. These encompass, among others, projectile
and fragment sizes, depths of intrusion, crater depths, and
fragment velocities.

The variable parameters in the experiment are the projectile
velocity, the impact angle, the physical agglomerate properties
(see Table \ref{targetprojectiles}) and the projectile mass. Fig.
\ref{experimentalparametersa} gives an overview of the projectile
masses, the impact velocities and the impact angles for the three
agglomerate compositions described in Table
\ref{targetprojectiles} and impacts into unprocessed targets. For
a better statistical representation, the impact angle $\theta$
(relative to the target normal) is replaced by the squared sine of
the above-defined angle, $\sin^2(\theta)$, so that each
$\sin^2(\theta)$-interval has the same statistical probability for
random collisions. The original projectiles had sizes of typically
1 mm and, thus, masses of $m_{\rm proj} \sim 10^{-6}$ kg. During
the release of these projectiles, a small fraction of their mass
fragmented off and in most cases also hit the target. This means
that we could observe, in addition to the original projectiles,
several collisions between projectiles of sub-mm sizes and the
target agglomerates. The lower limit of agglomerate masses of
$\sim 10^{-9}$ kg is due to the finite resolution of the cameras
(see above). The target masses are in the range $m_{\rm targ}
\approx 1.0 \ldots 1.5$ g so that $m_{\rm proj} \ll m_{\rm targ}$
for all impacts. Thus, our impact experiments are valid for all
target masses $m_{\rm targ} \gg m_{\rm proj}$ and are not
restricted to target sizes of centimeters. Impact velocities
ranged from $\sim 0.1~\rm m~s^{-1}$ to $\sim 3~\rm m~s^{-1}$, with
a slight systematic increase in velocity with increasing
projectile mass. This effect -- with a typical difference in
impact velocity between the largest and the smallest projectiles
of $\sim 0.5~\rm m~s^{-1}$ -- is due to some residual friction of
the travelling projectiles with the rarefied-gas atmosphere of
$\sim 20$ Pa pressure. The total number of microgravity
experiments performed for these experiments is 45. The experiments
were carried out in three series between November 2003 and October
2004.

Fig. \ref{experimentalparametersc} shows the parameter space of
the impact experiments of $\rm SiO_2$ dust aggregates with $\phi =
0.15$ into ``molded'' targets with $\phi = 0.15-0.20$. These data
were collected in a drop-tower campaign comprising 9 flights in
April 2006. Due to the morphology of the target surface (see Fig.
\ref{targetprojectiles}), the variation of the impact angle is
irrelevant so that all impacts were carried out normal to the
target surface.

\section{\label{sectexpres}Experimental Results}

\subsection{\label{stpr}Sticking Properties}
\begin{figure}[ht]
  \plotone{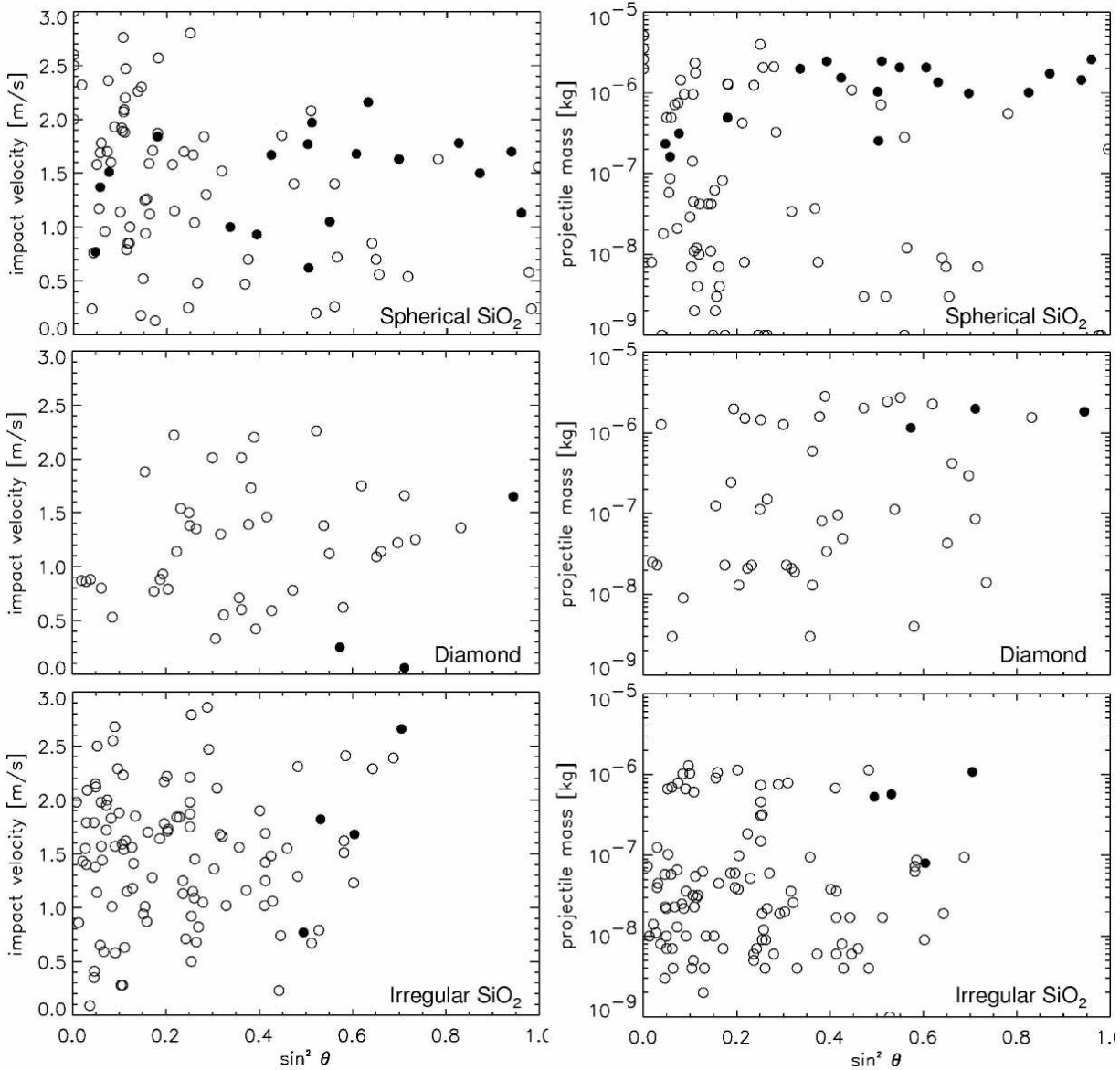}
  \figcaption{\label{experimentalparametersa}The distribution of
  the experiment parameters impact angle, projectile
  mass and impact velocity for the
  three agglomerate types consisting of spherical, monodisperse
  $\rm SiO_2$ particles (top), irregular quasi-monodisperse
  diamond particles (middle) and polydisperse, irregular
  $\rm SiO_2$ particles (bottom). Open symbols denote collisions which
  resulted in sticking, filled symbols those in which the projectile
  bounced off the target after the impact.}
\end{figure}

\begin{figure}[htp]
  \plotone{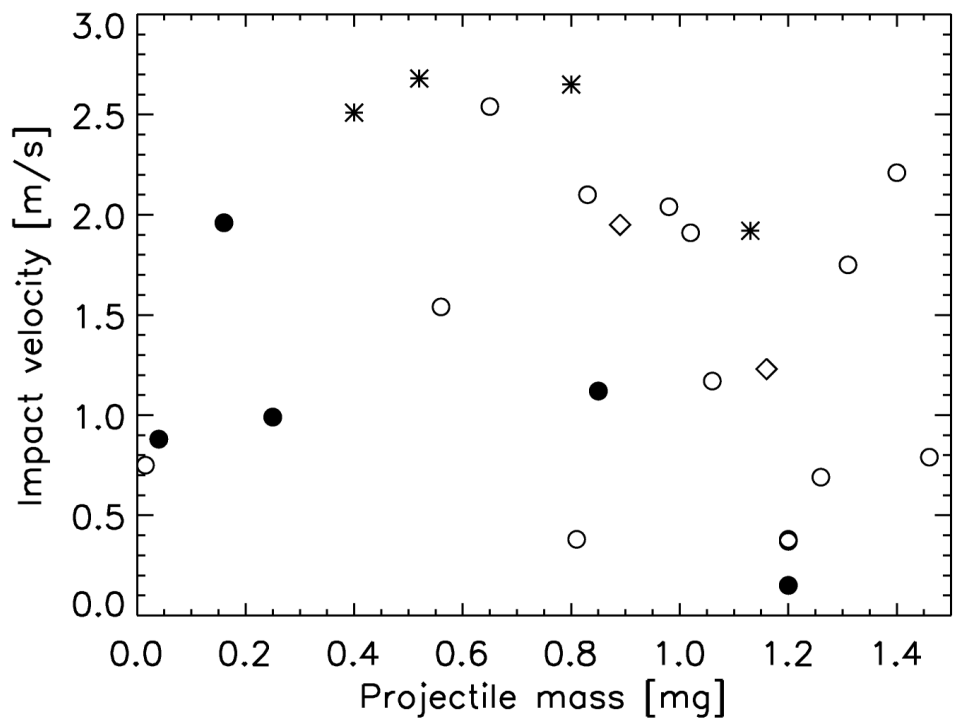}
  \figcaption{\label{experimentalparametersc}Masses and impact
  velocities of the agglomerates consisting of spherical,
  monodisperse $\rm SiO_2$ particles with $\phi = 0.15$ for
  impacts into ``molded'' targets with $\phi = 0.15-0.20$. The open circles denote sticking, the
  full circles denote bouncing, the asterisks denote fragmentation
  without mass transfer to the target, and the diamonds denote
  fragmentation with mass transfer to the target.}
\end{figure}

\begin{figure}[htp]
  \plotone{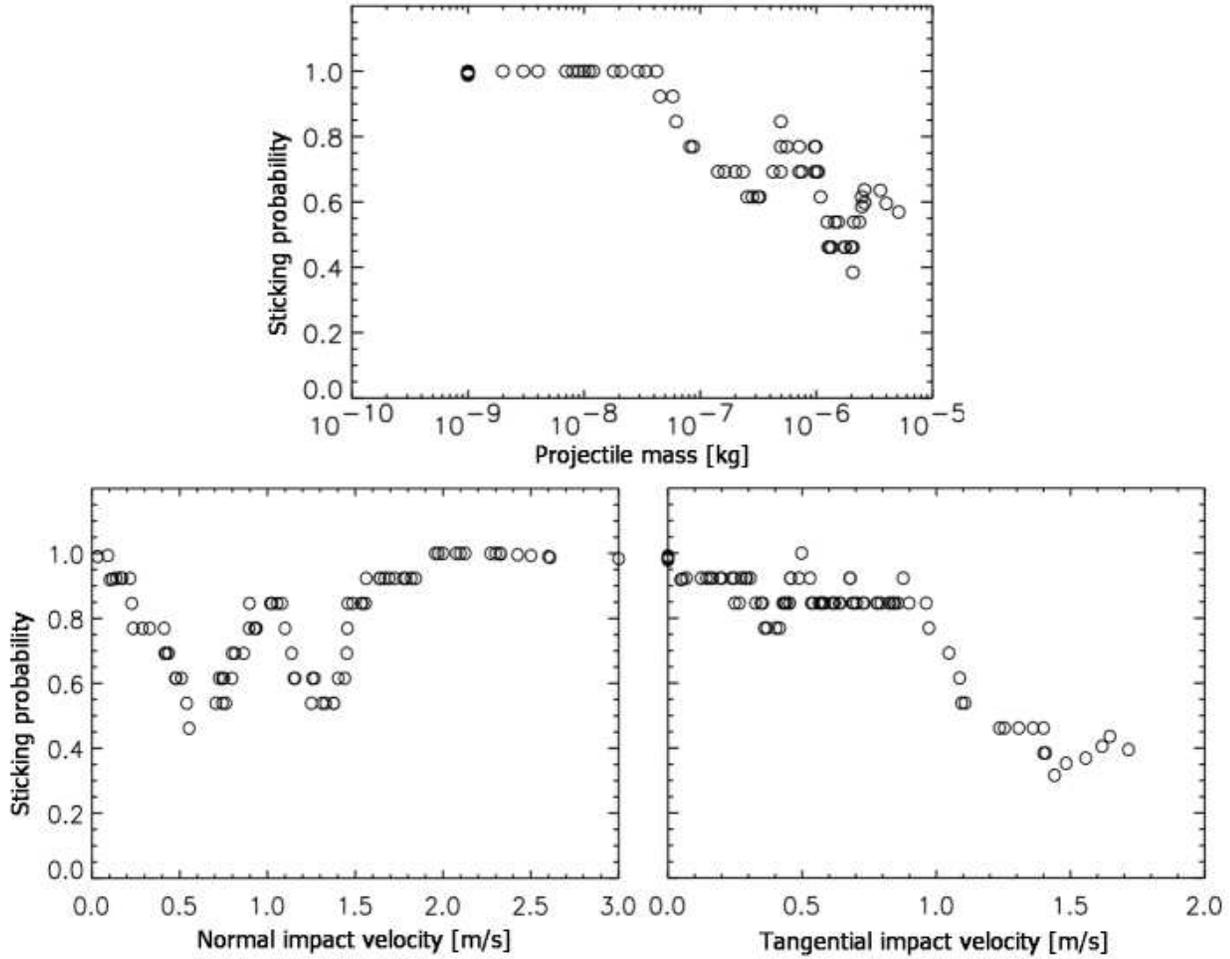}
  \figcaption{\label{stickingprobability} Sticking probability in collisions
  between dust-aggregate projectiles and targets consisting of
  $1.5~\rm \mu m$ $\rm SiO_2$ spheres with a volume filling factor
  of $\phi =0.15$. The sticking probabilities were derived by
  averaging over 13 impacts.}
\end{figure}

For dust aggregates consisting of irregular $\rm SiO_2$ and
diamond particles the sticking probability in the mass -- velocity
-- impact angle range is very close to unity (see Fig.
\ref{experimentalparametersa}), with only a few projectile
aggregates bouncing from the target. For dust aggregates
consisting of spherical, monodisperse $\rm SiO_2$ grains, i.e. for
the densest projectiles and targets, the distribution of the
sticking (open circles in Fig. \ref{experimentalparametersa}) and
non-sticking collision events (full circles in Fig.
\ref{experimentalparametersa}) in the parameter space is not
random. It is evident that the non-sticking collisions into the
flat targets occur preferentially for intermediate velocities of
$\sim 1-2~\rm m~s^{-1}$, for larger impact angles, and for more
massive projectiles. A determination of the sticking probability
as a function of projectile mass and the components of the impact
velocity normal and tangential to the target surface is shown in
Fig. \ref{stickingprobability}. Sticking probabilities were
derived by sliding averaging over 13 data points sorted in
projectile mass and normal/tangential impact velocity. From Figs.
\ref{experimentalparametersa} and \ref{stickingprobability} it is
clearly visible that (1) the sticking probability is $\beta = 1$
for aggregates with $m \lesssim 10^{-7}$ kg and falls steadily to
values $\beta \approx 0.5$ for the highest aggregate masses of $m
\approx 5\cdot10^{-6}$ kg, (2) for both, very low ($v \lesssim 0.5
\rm ~ m~s^{-1}$) and very high ($v \gtrsim 2 \rm ~ m~s^{-1}$)
normal impact velocities, the sticking probability is $\beta
\approx 1$ (however, mind that projectile velocity and mass are
not independent parameters and that the slowest impact velocities
stem from the smallest projectiles), whereas the sticking
probability drops to $\beta \approx 0.5$ for normal impact
velocities in the range $0.5 {\rm ~ m~s^{-1}} \lesssim v \lesssim
1.5 {\rm ~ m~s^{-1}}$, (3) the sticking probability decreases
steadily for increasing tangential impact velocity and reaches
values as low as $\beta \approx 0.4$ for the highest tangential
velocities of $\sim 1.7~\rm m~s^{-1}$.

For the ``molded'' and slightly compressed targets, sticking is
even the exception (see Fig. \ref{experimentalparametersc}), while
for the flat and fluffy targets most collisions result in mass
gain of the target. We find a sticking probability of $\beta
\approx 0.2$ when the target is slightly compacted and has local
radii of curvature comparable to the projectile radii. It was
observed that the projectiles stuck to the target only when they
accidentally hit a ``valley''.

\begin{figure}[ht]
  \plotone{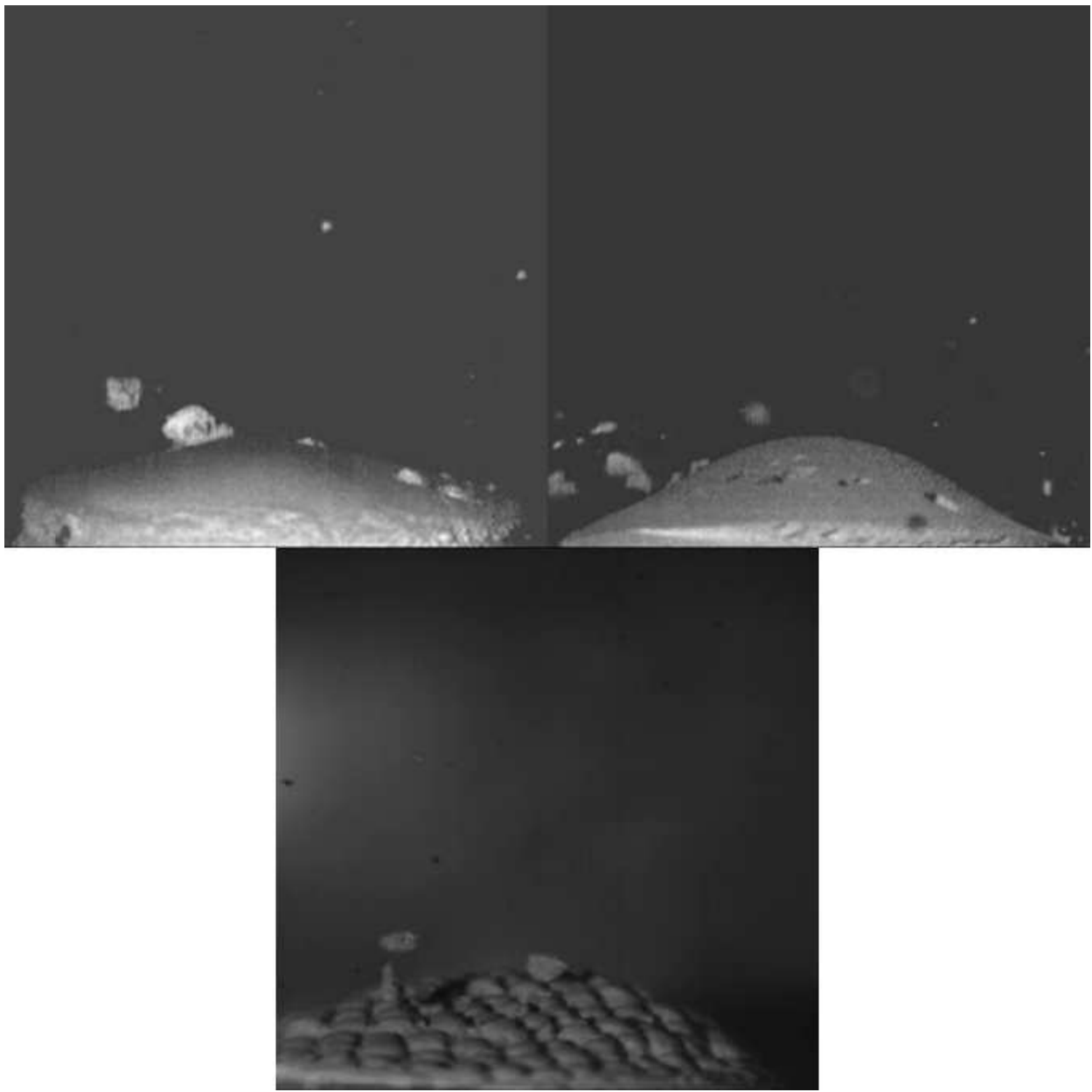}
  \figcaption{\label{movie_examples} Examples for collisions of
  high-porosity dust-aggregate projectiles into targets of the same
  composition. The width of the full image is 27.5 mm.
  Top left: irregular $\rm SiO_2$ particles; impact velocity
  $v \approx 0.8-1.0 ~\rm m~s^{-1}$. Top right: diamond particles;
  impact velocity $v \approx 1.7-2.3 ~\rm m~s^{-1}$. Bottom:
  spherical $\rm SiO_2$ particles; ``molded'' target; impact
  velocity $v \approx 0.38~\rm m~s^{-1}$. Movies are available
  in the online version of this paper.}
\end{figure}

Fig. \ref{movie_examples} shows example movies into unprocessed
and ``molded'' target aggregates (online version only).

\subsection{\label{sectmasstrans}Mass transfer in collisions}
Due to the low number of non-sticking collisions for the
agglomerates consisting of irregular monomers, we could not
perform a statistical analysis of the properties of the bouncing
agglomerates for non-spherical monomers. Thus, this and the
following subsections will mainly deal with the analysis of the
non-sticking collisions of agglomerates consisting of spherical,
monodisperse $\rm SiO_2$ particles into flat soft targets of the
same material.

\begin{figure}[ht]
  \plotone{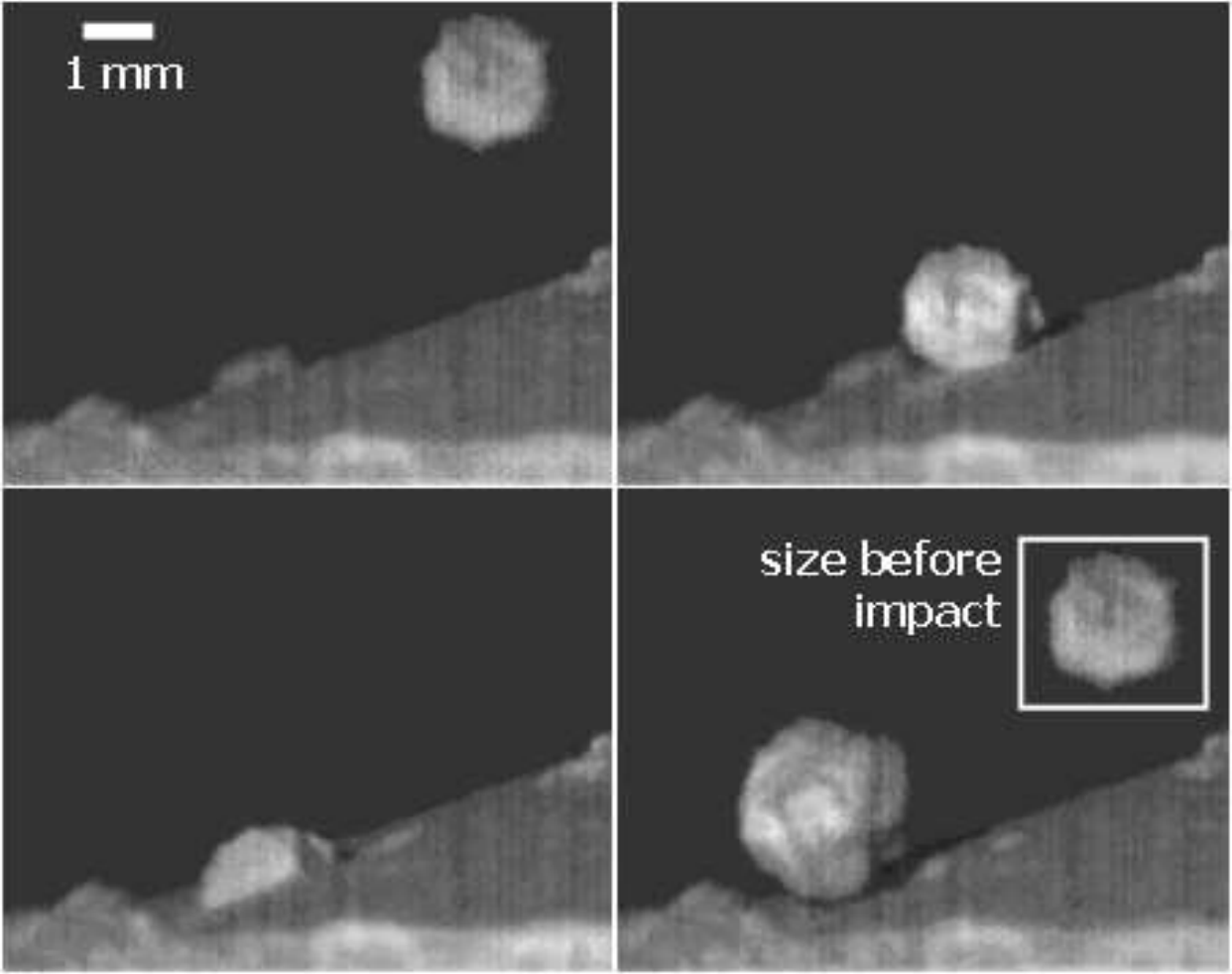}
  \figcaption{\label{masstransferimage}Example of a non-sticking
  collision with mass transfer
  between the target and the projectile agglomerate (the movie
  of this impact can be found in the online material to Fig.
  \ref{movie_examples}). Shown is a
  sequence of four images recorded with the high-resolution
  high-speed camera. The inset in the bottom right frame is a copy of
  the projectile from the first frame. Projectiles and target consist of
  1.5 $\rm \mu m$ $\rm SiO_2$ spheres and have volume filling factors
  of $\phi = 0.15$. The impact velocity is $v = 1.77~\rm m~s^{-1}$.}
\end{figure}

\begin{figure}[ht]
  \plotone{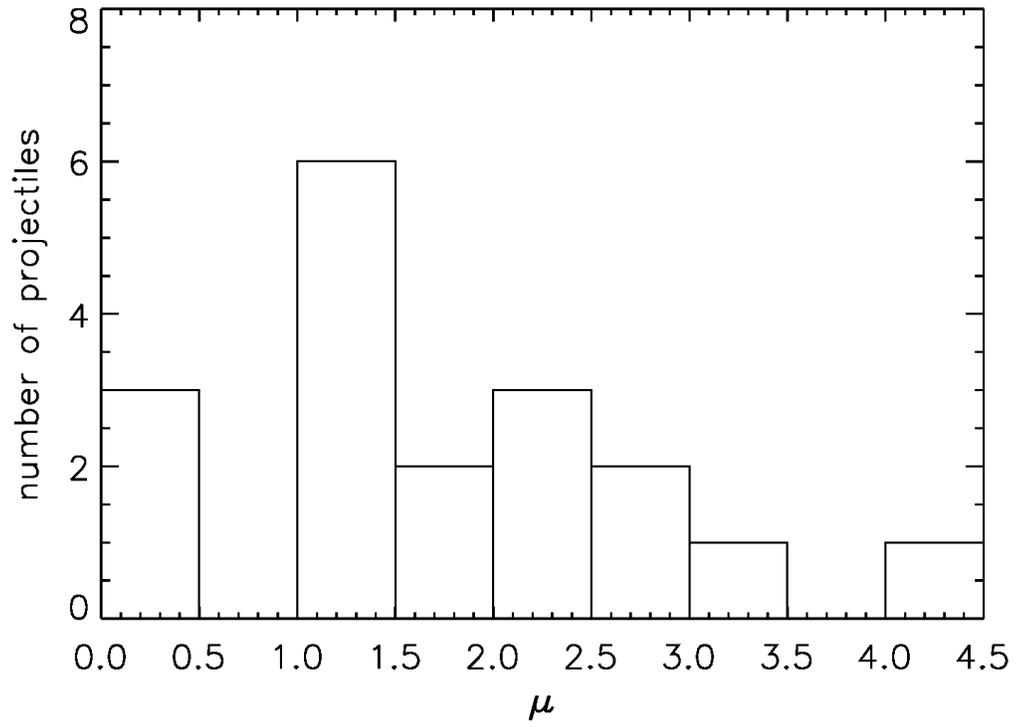}
  \figcaption{\label{masstransferhistogram}Histogram of the mass
  ratio $\mu$ defined in Eq. \ref{masstransfer}.}
\end{figure}

Fig. \ref{masstransferimage} displays an example of a collision in
which the impinging agglomerate bounced off after the collision.
From the comparison between the first and the last image of the
sequence (see inset) it is evident that the size of the projectile
agglomerate changed. We determined the mass ratio of the
projectile immediately after and before the impact by
\begin{equation}\label{masstransfer}
    \mu = \frac{m^{\prime}}{m} \approx
    \left(\frac{s^{\prime}_{\rm max} \cdot s^{\prime}_{\rm min}}
    {s_{\rm max} \cdot s_{\rm min}}\right)^{3/2} \, ,
\end{equation}
in which $s^{\prime}_{\rm max}$, $s^{\prime}_{\rm min}$, $s_{\rm
max}$, and $s_{\rm min}$ denote the maximum and minimum linear
extension of the agglomerate after and before the impact,
respectively. Fig. \ref{masstransferhistogram} shows the
distribution of the derived mass ratios. Although the statistics
is still somewhat poor, it can be seen that the mass transfer
between target and projectile agglomerate can be considerable and
obtains mostly values in the range $\mu = 1 \ldots 4$. The mean
mass ratio is $\bar{\mu}=2.1$. This means that on average a
non-sticking impact leads to a considerable mass loss of the (more
massive) target agglomerate, a process which has not been
considered before.

\subsection{\label{cor}Coefficient of restitution and energy loss}
\begin{figure}[htp]
  \plotone{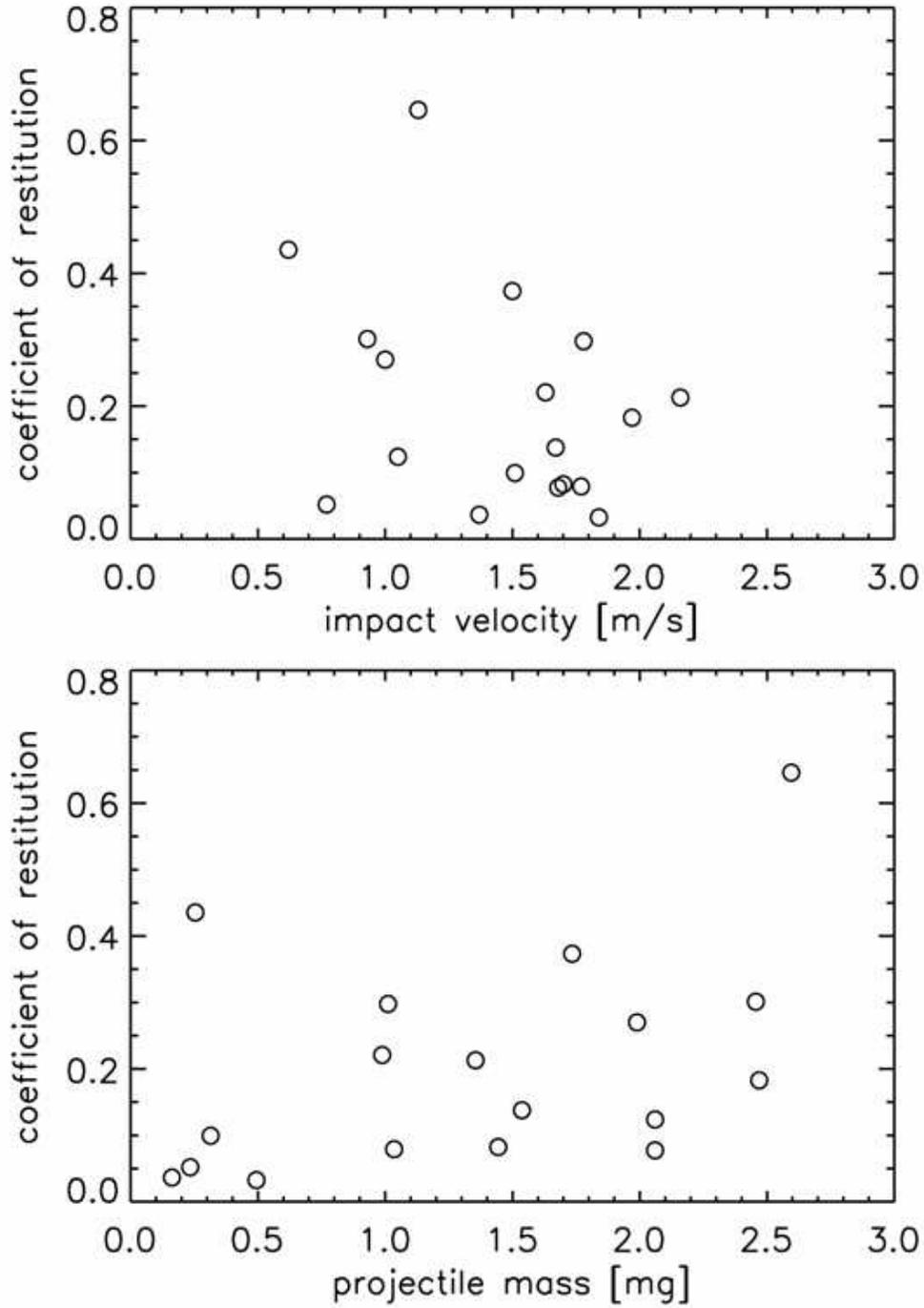}
  \figcaption{\label{coefficientofrestitutionfig}Coefficient of restitution
  as a function of impact velocity (top) and impactor mass
  (bottom) for the 18 non-sticking impacts of agglomerates consisting of
  $\rm SiO_2$ spheres. }
\end{figure}

In the case of non-sticking the kinetic energy is not fully
absorbed within the projectile and target agglomerates. A usual
method to describe the amount of plasticity in a collision is by
using the coefficient of restitution, defined by
\begin{equation}\label{coefficientofrestitution}
    \epsilon = \frac{v^{\prime}}{v} \, .
\end{equation}
Here $v^{\prime}$ and $v$ denote the relative velocity between the
projectile and the target after and before the collision. Fig.
\ref{coefficientofrestitutionfig} shows the coefficients of
restitution for all 18 non-sticking impacts of agglomerates
consisting of $\rm SiO_2$ spheres as a function of impact velocity
and projectile mass. The data points are scattered between
relative low $\epsilon < 0.1$ and rather high values $\epsilon >
0.4$ with no apparent dependence of the coefficient of restitution
on the impact velocity and on the projectile mass. The mean value
of the coefficient of restitution is $\bar{\epsilon} = 0.20$ and
the root mean square value is $\sqrt{\bar{\epsilon^2}} = 0.16$.

\begin{figure}[htp]
  \plotone{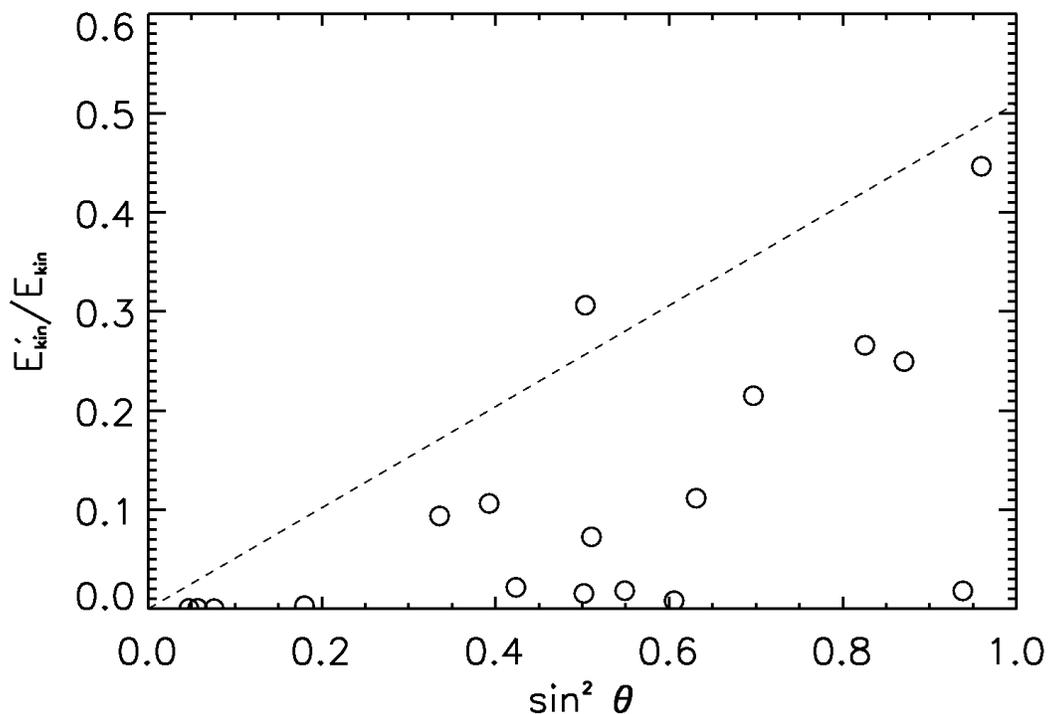}
  \figcaption{\label{energyloss}The ratio of total kinetic energy after and before the
  collision as a function of the squared impact parameter (circles). The
  straight line gives the upper limit derived by \citet{blumue93},
  which is valid for perfectly plastic central and perfectly
  elastic and non-slipping grazing collisions.
  The energy loss for central collisions
  ($\sin^2{\theta}=0$) in this model
  is due to plasticity and for grazing collisions ($\sin^2{\theta}=1$)
  is due to an energy
  transfer from translation to rotation.}
\end{figure}

The coefficient of restitution plays an important role in the
dense dust-dominated subdisk or inside condensations caused by the
streaming instability, in which mutual collisions among the dust
aggregates can act as ``cooling'' \citep{johetal07} . For the
physical processes inside colliding dust aggregates, it is,
however, more interesting to consider the ratio of total kinetic
energy after and before the collision (taking into account the
mass transfer from target to projectile) as a function of the
squared impact parameter. Fig. \ref{energyloss} shows this data.
Also plotted in Fig. \ref{energyloss} is the linear relation
\begin{equation}
\label{transrot}
    \frac{E^{\prime}_{\rm kin}}{E_{\rm kin}} = \epsilon^2(0) +
    \epsilon^2(1) \cdot \sin^2{\theta}
\end{equation}
which is for $\epsilon^2(0) = 0$ (perfectly inelastic central
collisions) and $\epsilon^2(1) = (5/7)^2 = 0.51$ (pure frictional
transition from translational to rotational motion and no
plasticity for glancing collisions) an upper limit for
dust-aggregate collisions \citep{blumue93}. It is evident that our
data fall much below the upper limit given by Eq. \ref{transrot}.
This means that plasticity does not only play a role for the
normal but also for the tangential component of the collision,
i.e. the shear strength of the aggregate material is overcome
during the impacts. However, the data in Fig. \ref{energyloss}
shows an increasing trend of $\frac{E^{\prime}_{\rm kin}}{E_{\rm
kin}}$ with increasing $\sin^2{\theta}$, which was also observed
by \citet{blumue93}, i.e. the total amount of plasticity decreases
with increasing impact parameter.

\subsection{Semi-elastic rebound and tensile strength}
\begin{figure}[ht]
  \plotone{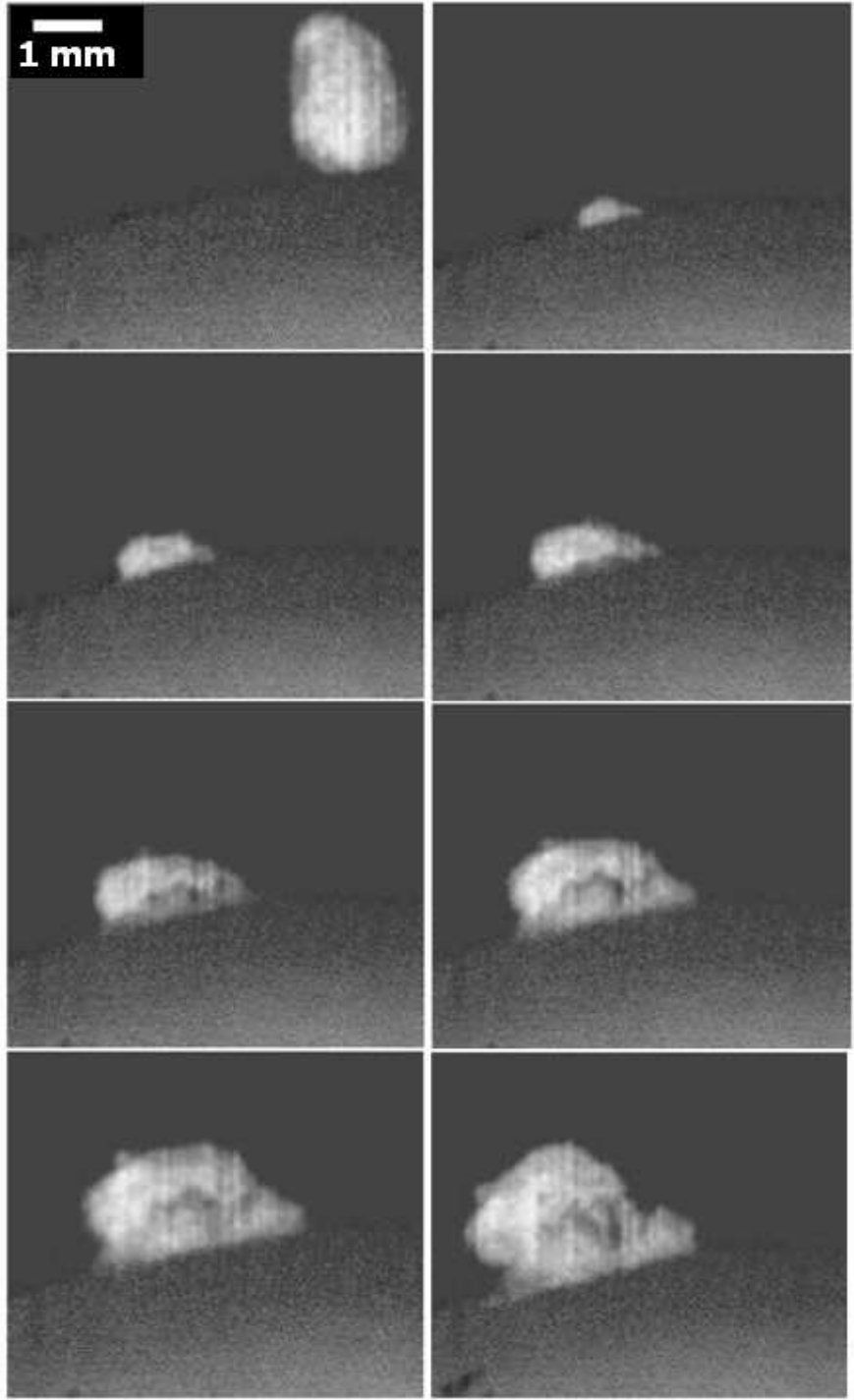}
  \figcaption{\label{semielasticrebound}An example of a
  partial rebound of a projectile agglomerate consisting of
  irregular $\rm SiO_2$ particles (the movie
  of this impact can be found in the online material to Fig.
  \ref{movie_examples}). On the top left
  image the projectile can be seen just before the impact. The impact
  angle is $\theta = 39.9^{\circ}$. The subsequent images -- taken 4.3, 8.6, 12.9,
  21.5, 34.4, 47.3, and 137.6 ms after the first image -- show that the
  projectile penetrates into the target and then is slowly rebounding
  with decreasing velocity until it finally sticks. The impact velocity is
  $v = 1.02~\rm m~s^{-1}$.}
\end{figure}

\begin{figure}[htp]
  \plotone{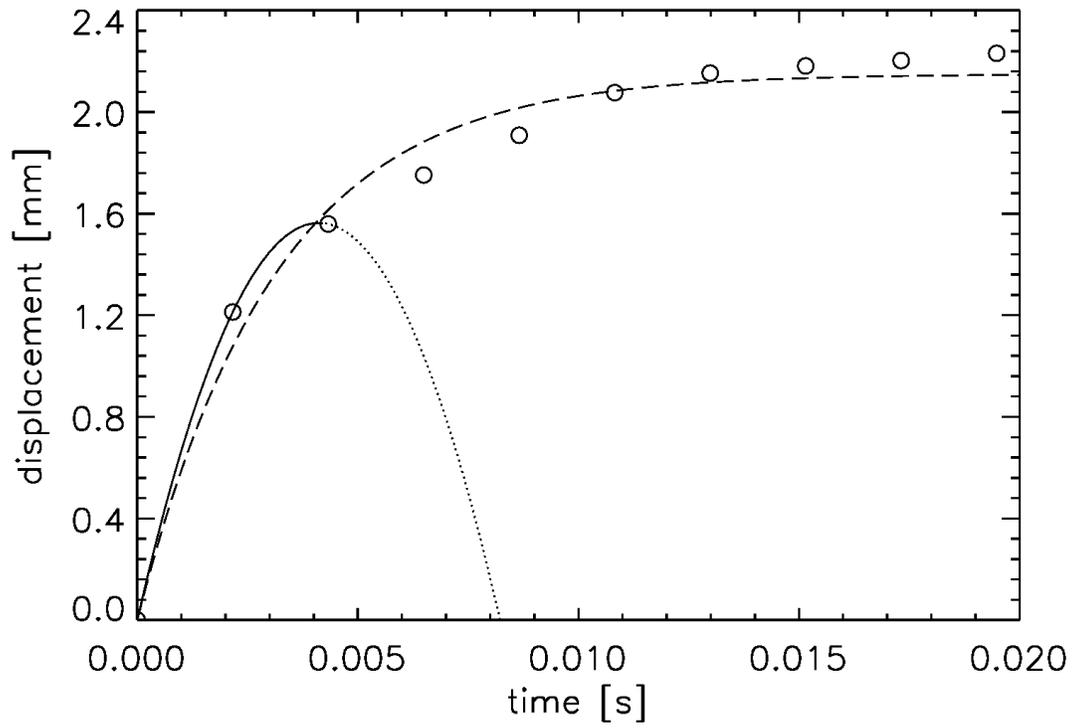}
  \figcaption{\label{reboundingparticle}Displacement of the rebounding projectile
  in Fig. \ref{semielasticrebound} as a function of time. Also shown are
  two fit functions following Eqs. \ref{parabola} and
  \ref{exponential}}.
\end{figure}

Even for the cases in which the projectile agglomerates stuck to
the target after the collision, the total kinetic energy was
initially not fully dissipated into plastic deformation of the
aggregates. An example is given in Fig. \ref{semielasticrebound}.
After impacting the target, the projectile (consisting of
irregular $\rm SiO_2$ particles) is rebounding but is not able to
escape from the target due to a too strong inter-particle
attraction. Fig. \ref{reboundingparticle} shows the temporal
dependence of the displacement of the projectile from its deepest
penetration during the rebounding phase of its trajectory.

The example in Fig. \ref{semielasticrebound} also demonstrates how
the mass transfer described in Sect. \ref{sectmasstrans} works: on
the way out of the target, the projectile gets decelerated (see
Fig. \ref{reboundingparticle}) by the adhesion forces between the
projectile and the target. As long as the adhesion is larger than
the tensile strength (see below), the projectile is attached to
the target and drags target material along. In the cases of
non-sticking, the tensile strength of the macroscopic dust
aggregate (see Table \ref{table1}) is overcome with some of the
target material sticking to the projectile agglomerate. It is
clear that the tensile strength is almost reached by the
rebounding particle in Fig. \ref{semielasticrebound}. Thus, we can
use the data in Fig. \ref{reboundingparticle} to estimate the
dynamic tensile strength of the high-porosity agglomerate and
compare these value to the static measurements of the same
material. We can fit a parabolic function
\begin{equation}\label{parabola}
    l(t) = \frac{1}{2} ~ a_0 ~ t^2 + v_0 ~t \, ,
\end{equation}
to the first three data points in Fig. \ref{reboundingparticle}
and get $a_0 = -185 \, \rm m~s^{-2}$ and $v_0 = 0.76 \, \rm
m~s^{-1}$ for the maximum initial acceleration and velocity of the
rebounding projectile (solid curve in Fig.
\ref{reboundingparticle}). In addition to that, we fit a decaying
exponential function of the form
\begin{equation}\label{exponential}
    l(t) = l_0 ~ \cdot ~ \left(1-\exp\left[-\frac{t}{\tau}\right]\right) \, ,
\end{equation}
with $l_0 = 2.2 \cdot 10^{-3}$ m and $\tau = 3.1 \cdot 10^{-3}$ s,
to all data points shown in Fig. \ref{reboundingparticle} (dashed
curve). Both functions are not motivated by physical
considerations but only help to estimate the initial acceleration.
Differentiations of the function in Eq. \ref{exponential} yield
the initial velocity and acceleration of $v_0 = 0.71 \, \rm
m~s^{-1}$ and $a_0 = -227 \, \rm m~s^{-2}$. With an estimated
cross section of the particle of $A = 1.5 \cdot 10^{-6} \, \rm
m^2$ and an estimated mass of the rebounding agglomerate of $m =
1.7 \cdot 10^{-6}$ kg (including the mass transfer from the target
to the impinging projectile), we get for the lower limit of the
tensile strength, $T \stackrel{>}{\sim} m a_0 / A$, values of
\begin{equation}
\label{tensile1}
    T \stackrel{>}{\sim} 210 \, \rm N~m^{-2}
\end{equation}
for the parabolic function and of
\begin{equation}
\label{tensile2}
    T \stackrel{>}{\sim} 260 \, \rm N~m^{-2}
\end{equation}
for the decaying exponential function, respectively. These values
are in good agreement with the static values  $T = 300 \, \rm
N~m^{-2}$ (see paper I and Table \ref{table1}).

\subsection{Intrusion and crater depths}
\begin{figure}[ht]
  \plotone{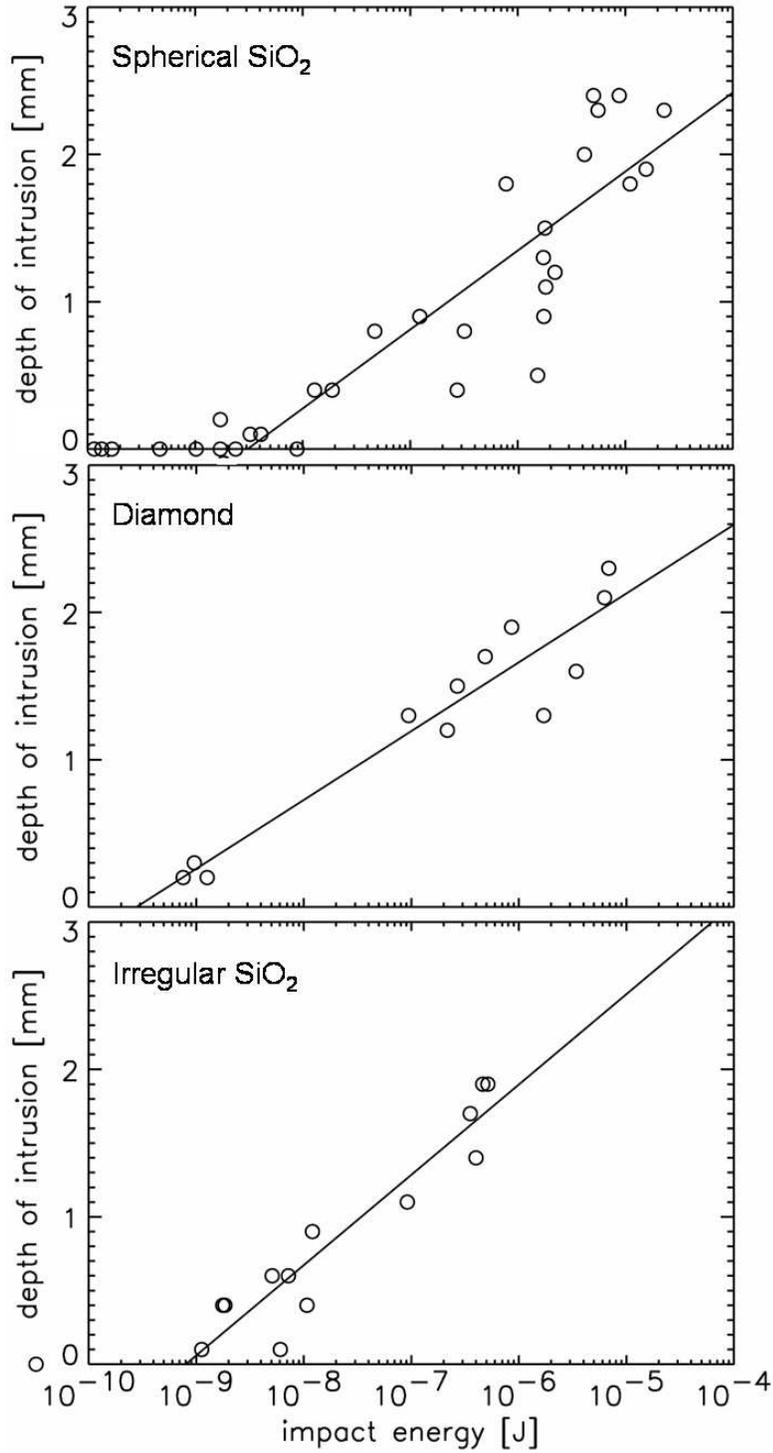}
  \figcaption{\label{doienergy}Depth of intrusion for all sticking events as a function
  of the total kinetic energy of the projectiles
  for all three agglomerate types. The
  least-squares fits following Eq. \ref{intrusion2} are indicated by the
  three lines.}
\end{figure}

For those impact events in which the projectiles stuck to the
target agglomerate we could determine the depth of intrusion
$d_{\rm i}$ perpendicular to the surface. It turned out that the
depth of intrusion was generally larger for higher impact
energies. Fig. \ref{doienergy} shows the data for all projectile
and target materials as a function of the kinetic impact energy of
the projectiles.

In order to penetrate into the target agglomerate, the projectiles
need to overcome the compressive strength of the target. The data
in Fig. \ref{doienergy} suggest that a threshold energy $E_{\rm
min}$ is required to yield a finite penetration depth. For the
determination of this value, we assume a linear dependence between
the depth of intrusion and the logarithm of the impact energy, as
suggested by the data in Fig. \ref{doienergy}. A least squares fit
of
\begin{equation}\label{intrusion2}
    d_{\rm i} = x \log\left(\frac{E_{\rm kin}}{E_{\rm
    min}}\right)\,
\end{equation}
(for $E_{\rm kin} \ge E_{\rm min}$) results in $x = 0.54~\rm mm$
and $E_{\rm min} = 3.1 \cdot 10^{-9} ~ \rm J$ for spherical $\rm
SiO_2$, $x = 0.47~\rm mm$ and $E_{\rm min} = 2.8 \cdot 10^{-10} ~
\rm J$ for irregular diamond, and $x = 0.61~\rm mm$ and $E_{\rm
min} = 8.0 \cdot 10^{-10} ~ \rm J$ for irregular $\rm SiO_2$,
respectively (solid lines in Fig. \ref{doienergy}).

If we only consider the projectiles whose energies are close to
the threshold energy $E_{\rm min}$, we find that their masses are
$m \approx 10^{-9}$ kg. With aggregate densities of 300 $\rm
kg~m^{-3}$, 390 $\rm kg~m^{-3}$, and 182 $\rm kg~m^{-3}$ for
spherical $\rm SiO_2$, diamond, and irregular $\rm SiO_2$, we get
typical projectile volumes at the onset of intrusion of $V = 3.3
\times 10^{-12}~\rm m^3$, $V = 2.6 \times 10^{-12}~\rm m^3$, and
$V = 5.5 \times 10^{-12}~\rm m^3$, respectively. Using the
above-derived minimum impact energies for intrusion, $E_{\rm
min}$, we can derive the critical impact pressure $p_{\rm min}
=E_{\rm min} / V = 940$ Pa, $p_{\rm min}=100$ Pa, and $p_{\rm
min}=150$ Pa, respectively, which is remarkably close to the
static compressive strengths of the respective target aggregates
(see Table \ref{table1}). Thus, for the onset of penetration, a
minimum impact pressure equivalent to the compressive strength is
required.

In addition to the depth of intrusion, we also derived for the
non-sticking events the crater depths. This was either done by a
determination of the shadow lengths inside the craters for oblique
illumination or by a measurement of the length of the escaping
projectile agglomerate. The crater depth is then approximated by
the difference of the length of the escaping projectile (which is,
due to the mass transfer between target and projectile, larger
than before the impact; see Sect. \ref{sectmasstrans}) and the
length of the part of the projectile protruding out of the target
at closest approach. A comparison of the crater depth and the
depth of intrusion for the case of spherical $\rm SiO_2$ monomers
(only in that case we have sufficiently many rebounding
projectiles) shows that the crater depths are a factor $\sim 3$
larger than the depths of intrusion. However, the mass transfer is
less than a factor of $\sim 3$ due to only a partial penetration
of the projectiles.

\subsection{Impact fragmentation and the
Influence of Local Surface Curvature and Density Enhancement} In a
few cases, the impinging projectile agglomerate fragmented upon
impact. Part of the residual fragments left the target
agglomerate, the other fragments stuck to the target. Most
fragmentation events occurred when the projectile either hit a
previously captured projectile or another surface irregularity
with low radius of curvature. Thus, we systematically investigated
the influence of the local radius of curvature of the target
aggregate on the outcome of a collision. For this, we used
previously ``molded'' target aggregates (see Sect. \ref{sectsetup}
and Fig. \ref{targetprojectiles}c) and made a series of impacts of
mm-sized, high-porosity projectile aggregates.

As already mentioned in Sect. \ref{stpr} and visible in Fig.
\ref{experimentalparametersc}, the sticking probability is
considerably reduced by the surface sculpting. Only 5 out of 25
projectiles stuck to the target, and sticking was apparently
restricted to those cases in which the projectile hit a ``valley''
on the target. In contrast to that, the fragmentation efficiency,
while negligible for impacts into smooth flat targets, increases
to 24\% (6 projectiles). However, most of the projectiles (14)
were semi-elastically rebounding after the impact, which makes the
rebound probability with 56\% the most likely single collisional
outcome for the impacts into ``molded'' targets.

It is clear from these experiments that, besides impact velocity,
impact angle, aggregate mass and aggregate packing density, the
local radius of curvature and, thus, the collisional history of
protoplanetary dust aggregates plays an important role for the
outcome of mutual collisions.

\section{\label{sectdisc}Discussion: Impact Dynamics of
High-Porosity Dust Agglomerates} From the results presented in the
previous section, we can derive some fundamental (although
approximate and preliminary) dynamical properties of high-porosity
dust aggregates. If we assume for simplicity that the aggregates
are fully elastic during collisions, we can use the well-known
Hertzian equation of motion \citep{hertz1882} for the collision
between a spherical particle of radius $s$ and mass $m$ and an
infinitely large sphere
\begin{equation}\label{hertz1}
    m \frac{{\rm d}^2 \delta}{{\rm d} t^2} + \frac{4}{3} s^{1/2}
    E^* \delta^{3/2},
\end{equation}
with $E^*$ being the elasticity parameter, defined by
\begin{equation}\label{hertz2}
    E^* = \frac{1}{2} \frac{E}{1 - \nu^2}
\end{equation}
for like materials. Here, $E$ and $\nu$ are the modulus of
elasticity and Poisson number of the materials, respectively. As
our aggregates very likely have Poisson numbers $\nu \approx 0$
(i.e. they are highly compressive), Eq. \ref{hertz2} reduces to
\begin{equation}\label{hertz3}
    E^* = \frac{1}{2} E .
\end{equation}

\begin{figure}[htp]
  \plotone{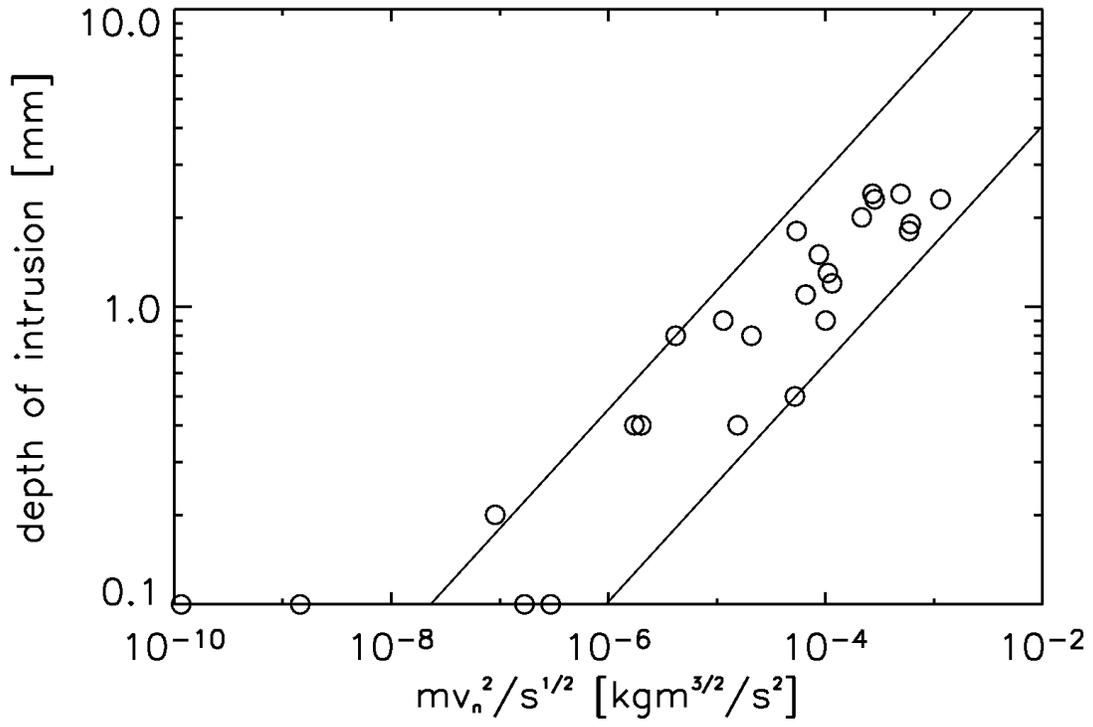}
  \figcaption{\label{hertzplot}Depth of intrusion of dust agglomerates consisting of spherical
  monodisperse $\rm SiO_2$ as a function of the parameter $\frac{m v_{\rm n}^2}{s^{1/2}}$.
  The indicated straight lines have slopes of 2/5 (see Eq. \ref{hertz4}).}
\end{figure}

With Eq. \ref{hertz3}, Eq. \ref{hertz1} can be solved for the
maximum penetration
\begin{equation}\label{hertz4}
    \delta_{\rm max} = \left( \frac{15 m v^2}{8 s^{1/2} E} \right)^{2/5} ,
\end{equation}
when $v$ is the initial collision velocity. If we identify
$\delta_{\rm max}$ with the measured intrusion of the projectiles,
$d_{\rm i}$, and plot these double-logarithmically as a function
of the parameter $\frac{m v_{\rm n}^2}{s^{1/2}}$, motivated by the
functionality in Eq. \ref{hertz4}, we see that we can indeed find
a slope close to 2/5 as predicted by Eq. \ref{hertz4} (Fig.
\ref{hertzplot}).

The two straight lines in Fig. \ref{hertzplot} formally give
$E=430$ Pa and $E=17,800$ Pa. Taking the geometric mean of these
two values, the agglomerates penetrate as deep as if they had a
modulus of elasticity of $\sim 2,800$ Pa. As the collisions are
clearly dominated by plasticity, we expect the compressive
strength to be of the same order as the modulus of elasticity,
i.e. $C \approx 2,800$ Pa. A comparison with Fig. 4 from paper I
shows that such compressions should lead to a compaction of $\phi
\approx 0.2$. Mind, however, that $E$ in Eq. \ref{hertz4} needs
not to be constant. As was shown by \citet{blusch04}, the
compressive stress $p_{\rm c}$ and the volume filling factor
$\phi$ are related through $p_{\rm c} \propto (\phi -
\phi_0)^{\beta}$ for $\phi_0 \le \phi \lesssim 0.22$, with $\phi_0
= 0.15$ and $\beta = 0.8$. Thus, we expect a similar relation
between $E$ and $\phi$ to be existent.

\begin{figure}[ht]
  \plotone{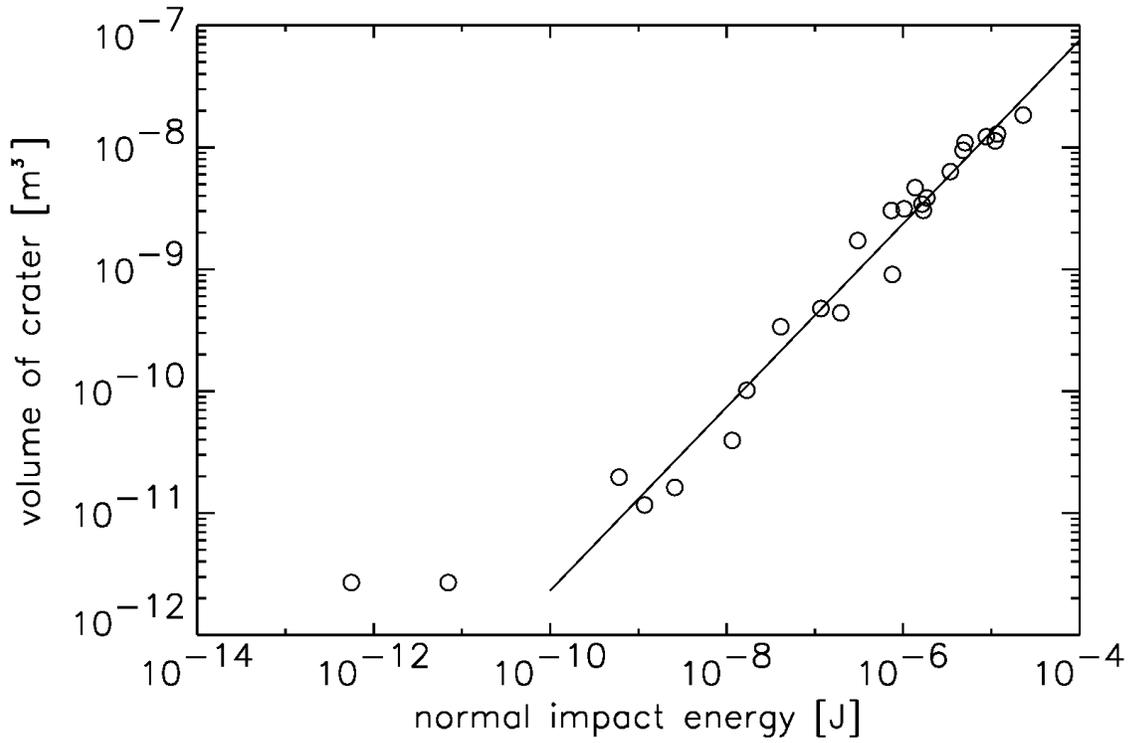}
  \figcaption{\label{cvenergy}Crater volume as a function of the
  normal component of the impact energy for agglomerates
  consisting of spherical $\rm SiO_2$. A power-law least squares
  fit to the data points $E > E_{\rm kin,n}$
  with a slope of 0.75 is plotted as a solid line.}
\end{figure}

\begin{figure}[ht]
  \plotone{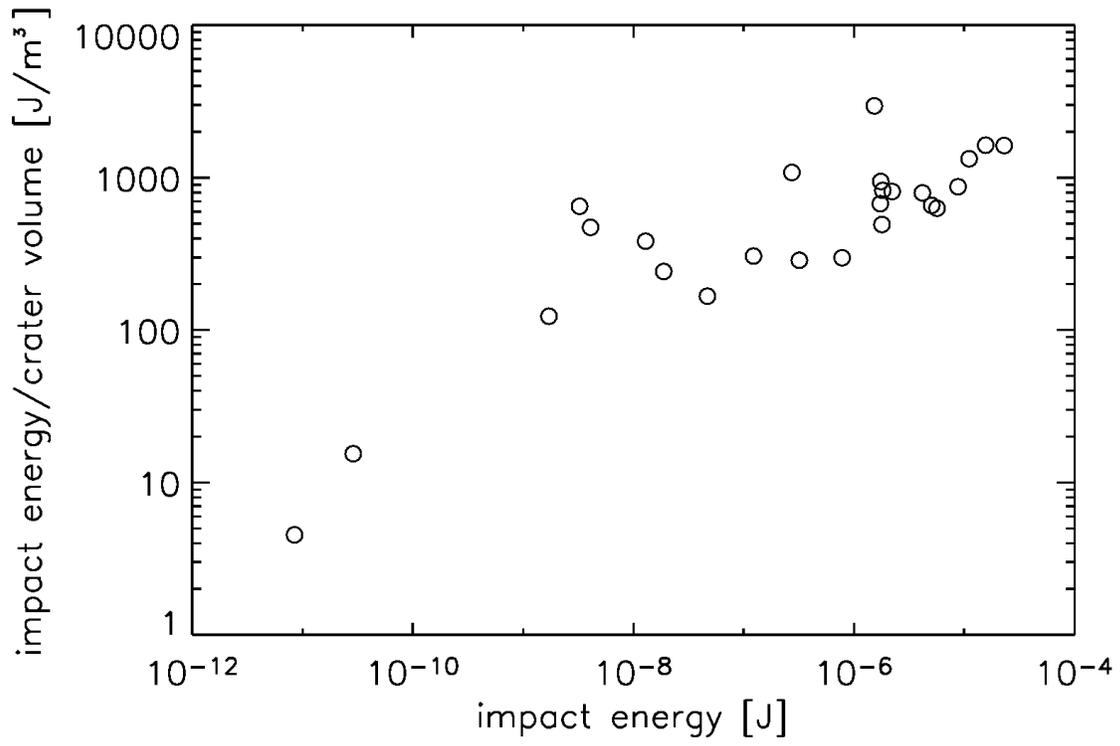}
  \figcaption{\label{pressureenergy}Dynamic pressure as a function of energy for agglomerates
  consisting of spherical $\rm SiO_2$.}
\end{figure}

If we define the crater volume by
\begin{equation}
    V = p ~ d_{\rm i}^2 ~ (s - \frac{d_{\rm i}}{3})
    \label{cratervolume1}
\end{equation}
for $d_{\rm i}<s$ and
\begin{equation}
    V = p ~ s^2  ~ d_{\rm i} ~ - \frac{\pi}{3} ~ s^3
    \label{cratervolume}
\end{equation}
for $d_{\rm i}\geq s$, with $s$ being the radius of the
projectile, we can approximately derive the dynamic impact
pressure $p_{\rm dyn}$ for fully plastic collisions
\citep{johnson1985}, defined by
\begin{equation}
    \label{impactpressure}
    p_{\rm dyn} = \frac{E_{\rm kin}}{V} .
\end{equation}
In Fig. \ref{cvenergy}, we plotted the crater volume as a function
of the normal component of the impact energy, $E_{\rm kin,n}$. It
is evident that there is a strong correlation between these two
quantities. We fit a power-law $V \propto E_{\rm kin,n}^{\alpha}$
to the data in Fig. \ref{cvenergy}, with $\alpha = 0.75 \pm 0.03$.
Thus, the dynamic impact pressure slightly increases with impact
energy from $\sim 200$ Pa to $\sim 2,000$ Pa for the impact energy
range between $\sim 3 \times 10^{-9}$ J and $\sim 2 \times
10^{-5}$ J (see Fig. \ref{pressureenergy}). The slightly
increasing values of $p_{\rm dyn}$ towards larger impact energies
is probably caused by the compaction of the aggregate volume
during impact. For the minimum impact energy for which intrusion
was found (see above), $E_{\rm min} = 3.1 \cdot 10^{-9} ~ \rm J$,
a minimum dynamic pressure of $\sim 300$ Pa is required to cause
the formation of a crater and, thus, compaction of the agglomerate
material in the contact zone between projectile and target
agglomerate. This pressure is very close to the minimum stress
under which the material yields (see Fig. 4 in paper I).

With this data, we can also explain the mass transfer observed in
many non-sticking collisions (see Sect. \ref{sectmasstrans} and
Figs. \ref{masstransferimage} and \ref{masstransferhistogram}).
During the impact, the projectile and part of the target volume
are slightly compacted. As was shown in paper I, aggregate
compaction increases the tensile strength of the material. Thus,
the compacted parts in the collisional volume have higher inner
cohesion so that they are more likely to keep sticking together
after the impact.

\section{\label{sectsummary}Summary}
We performed oblique impact experiments of typically 0.2-3 mm
diameter projectile agglomerates into 2.5 cm diameter target
agglomerates of identical composition. We used three different
particle types (monodisperse spherical $\rm SiO_2$ with 1.5 $\rm
\mu m$ diameter, irregular diamond with 1-2 $\rm \mu m$ diameter,
irregular $\rm SiO_2$ with 0.1-10 $\rm \mu m$ diameter) and
produced agglomerates by the random ballistic deposition process
\citep{blusch04}. Depending on the particle type, the agglomerates
had volume filling factors of $\phi = 0.15$ (for monodisperse $\rm
SiO_2$), $\phi = 0.11$ (for diamond), and $\phi = 0.07$ (for
irregular $\rm SiO_2$), respectively. Impact velocities ranged
between 0.1 and 3 m/s.

From the results of our experimental investigation presented in
the previous sections, we can draw the following conclusions:
\begin{enumerate}
    \item For a similar distribution of projectile masses, impact
    velocities and impact angles, the volume filling factor (porosity)
    of projectile and target has a considerable influence on the
    collisional outcome. Whereas very porous dust aggregates
    ($\phi \lesssim 0.10$) almost always stick, the sticking
    probability decreases with increasing filling factor for $\phi
    \gtrsim 0.15$ (Figs. \ref{experimentalparametersa} -
    \ref{stickingprobability}).

    \item For dust aggregates with volume filling factors
    $\phi \gtrsim 0.15$, projectiles with diameters $\gtrsim 1$ mm do only
    stick in near-central collisions, while near-grazing impacts
    lead to a rebound of the projectile (see Fig.
    \ref{experimentalparametersa}). The reason for the mass
    dependence of the sticking behavior
    is the interplay between the energy of the bouncing
    aggregates (which is, due to an almost mass-independent
    coefficient of restitution (see Fig.
    \ref{coefficientofrestitutionfig}), proportional to the aggregate mass)
    and the contact energy (which is roughly proportional to the
    aggregates' cross section). For small aggregates, the inertial
    energy of the rebounding aggregates is not capable of breaking
    the contacts, whereas large aggregates easily bounce off. For
    large aggregates, central collisions lead to a deeper
    penetration of the projectile aggregate into the target and
    thus to a larger contact area with stronger binding forces.

    \item Collisions which do not lead to sticking between projectile
    and target agglomerate result, on average, in a mass transfer
    from the larger to the smaller collision partner, i.e. to a
    mass loss of the target agglomerate (see
    Figs. \ref{masstransferimage} and \ref{masstransferhistogram}). This can be explained by an
    impact compaction of material comprising the projectile and
    part of the target and a resulting higher cohesion (tensile strength) of the
    compacted material.

    \item In the case of non-sticking, most of the kinetic energy
    of the projectile is dissipated in the collision. The residual
    energies of the bouncing projectiles (including the
    mass-transfer effect) are very small for near-central collisions
    and increase with increasing impact
    angle up to a few ten percent of the pre-collision energy
    (see Fig. \ref{energyloss}).

    \item The tensile strength of the agglomerates made of
    irregular $\rm SiO_2$ particles, as estimated in one
    experiment in which the projectile almost rebound after a
    collision (see Fig. \ref{semielasticrebound}), $T \gtrsim 260$ Pa,
    is in good agreement with the static measurements
    (see Table \ref{table1}). The tensile strength is responsible
    for aggregate cohesion and sticking upon impact.

    \item If we identify the compressive strength of the agglomerates
    made of spherical $\rm SiO_2$ particles with the estimated
    dynamic impact pressure (see Eq. \ref{impactpressure}
    and Fig. \ref{pressureenergy}), we see that the compressive
    strength increases with impact energy, as was already
    indicated by the static measurements in paper I. For impact
    velocities $\lesssim 3\, \rm m s^{-1}$ and projectile sizes
    $\sim 1$ mm, we get a maximum compressive strength of $C
    \approx 2,000$ Pa.

    \item Above a threshold value for the impact energy, $E_{\rm min}
    \approx 3\cdot 10^{-10} \ldots 3\cdot 10^{-9}$ J,
    a crater is formed. The crater volume scales with the normal
    component of the impact
    energy as $V \propto E_{\rm kin,n}^{0.75}$.
    The corresponding
    dynamical pressure, $p_{\rm dyn} = E_{\rm kin}/V$ is
    only weakly dependent on impact energy, $p_{\rm dyn} \propto
    E_{\rm kin}^{0.25}$ and ranges from $\sim 200$ Pa to
    $\sim 2,000$ Pa in our
    experiments. The threshold value for crater formation of
    $p_{\rm dyn} \approx 200$ Pa corresponds well with the onset of
    compaction measured by \citet{blusch04} and paper I.
    The largest
    crater volumes were found for the highest impact energies
    close to the transition between sticking and non-sticking.
    These crater volumes (see Fig. \ref{cvenergy}) are typically
    $V = 10^{-8} ~ \rm m^3$, Comparison with the masses of the
    largest (and, thus, most energetic) projectile agglomerates
    and the projectile
    densities (see data in Table \ref{table1}) show that the
    corresponding projectile volumes are also $V_{\rm proj}
    \approx 10^{-8} ~ \rm m^3$. Thus, the amount of compaction
    of the target and projectile agglomerates in these cases is
    not negligible. From \citet{blusch04} we can see that at
    a compression of $2,000$ Pa, the volume filling
    factor increases from $\phi = 0.15$ to $\phi = 0.20$.
    If these static compaction
    measurements are transferrable to the dynamic problem,
    we conclude that a volume
    at least as large as twice the projectile's
    is compressed to $0.20/0.15 = 1.33$
    of its previous volume filling factor.

    \item Impact fragmentation was only rarely observed for impacts
    into flat and high-porosity targets. When the target surface
    was artificially roughened with a local radius of curvature
    of 1 mm (which resulted in a slight compaction of the target
    agglomerate to $\phi \approx 0.2$), the sticking
    probability was considerably reduced,
    and bouncing and fragmentation were the dominating processes.
    Earlier experiments by \citet{blumue93} also found impact
    fragmentation when two similar-sized dust aggregates collide at
    velocities of a few m/s. Thus, the local radius of
    curvature of two colliding dust aggregates plays a dominating
    role in the outcome of the collision.
\end{enumerate}

\section{\label{sectconclusion}Conclusions}
In paper I, we have shown that macroscopic protoplanetary dust
aggregates are expected to be very porous. Depending on the
collisional history and the size of the aggregates, we expect the
dusty objects to have volume filling factors between $\phi \approx
0.1$ (for all objects whose collision velocities never exceeded
$\sim 1~\rm m s^{-1}$, i.e. for sizes $\lesssim$ cm) and $\phi
\approx 0.3$ (for all objects with sizes $\gtrsim 1$ m). In this
paper, we confirmed that collisions among fluffy protoplanetary
dust aggregates with velocities $\sim 1~\rm m s^{-1}$ lead to
impact pressures of $\lesssim 2,000$ Pa and, thus, to a moderate
increase in volume filling factor. Collisions between
(sub-)mm-sized dust aggregates and cm- to dm-sized fluffy objects
are very abundant in the solar nebula and the impact velocities
range around $1~\rm m s^{-1}$ \citep{weicuz93}. Thus, our
microgravity experiments match the solar-nebula conditions very
closely and are directly applicable to growth models.

We find that the sub-mm sized dust aggregates always stick to much
larger target aggregates, independent of impact velocity and
impact angle. The larger, mm-sized aggregates, however, behave
differently. Sticking is restricted to the higher velocities and
to the smaller impact angles. This trend was found for dust
aggregates consisting of monodisperse spherical particles, for
quasi-monodisperse irregular particles as well as for irregular
monomer particles with a wide size distribution, although
quantitative differences in the sticking probabilities exist.
Thus, for random impacts in the solar nebula, the sticking
probability in collisions between mm-sized dust aggregates and
cm-dm sized dusty bodies is below unity. Moreover, non-sticking
(and even fragmentation) is favored when the local radius of
curvature at the point of impact of the larger body is similar to
the projectile radius. This can be the case for collisions between
similar-size dust aggregates or between projectiles and targets
with irregular, non-flat surface textures. Another interesting
feature we found in our experiments is that non-sticking impacts
(i.e. those with higher impact angles) lead to a mass transfer
from the larger to the smaller body.

As a consequence, future growth models for protoplanetary dust
should take into account that the outcome of a single collision
between dust aggregate A and dust aggregate B is not only
dependent on the size of A and B and the mutual collision
velocity, but is also influenced by the collisional history (e.g.
the distribution of local radii of curvature on the surfaces of A
and B; compaction) and the (random) impact angle of the particular
collision. In addition to that, due to the possible occurrence of
mass transfer from a larger body A to a smaller body B (in
bouncing collisions) or fragmentation of the smaller object B
(e.g. when it hits a surface part of A with a small local radius
of curvature), the numerical description of the collisional
outcome (and its use in e.g. Smoluchowski's growth equation)
becomes rather complex.

From our experimental findings, we consider it rather unlikely
that protoplanetary bodies can grow beyond dm-sizes in a direct
and simple hit-and-stick manner. Although our experiments suggest
that small projectiles stick at higher velocities than large
projectiles, this trend does not imply that km-sized bodies can
form by the accumulation of very small dust aggregates or single
particles. Recent experiments by Schr\"apler \& Blum (unpublished)
show that at impact velocities $\gtrsim 15~\rm m~s^{-1}$ dust
aggregates cannot grow by the accumulation of single grains (and,
therefore, also not by the accumulation of small dust aggregates).
However, m-sized bodies in protoplanetary disks possess relative
velocities with respect to small grains of the order of $50~\rm
m~s^{-1}$ \citep{weicuz93}. Moreover, experiments by
\citet{wuretal05a} suggest that impacts between dusty projectiles
into fluffy dusty targets around $\sim 10~\rm m s^{-1}$ never lead
to an accumulation of mass on the larger body but lead to strong
fragmentation of the projectile and cratering (i.e. mass loss) of
the target. A possible way out of this dilemma could be the
indirect effect of projectile or fragment capturing by aerodynamic
\citep{wuretal01a,wuretal01b} or electrostatic \citep{blu04}
forces or by gravitational collapse in locally overdense regions
in the midplane of protoplanetary disks \citep{johetal07}.

\acknowledgments
We are indebted to the German Space Agency DLR
for supporting this work (grant no. 50 WM 0336) and providing us
with the drop tower flights. We thank the staff at the ZARM drop
tower facility for their help and hospitality during our
campaigns.

\setlength{\hoffset}{-7mm}
\begin{deluxetable}{lcccl}
\tabletypesize{\footnotesize}
\tablecaption{\label{table1}Physical parameters of the spherical
$\rm SiO_2$ particles, the diamond grains, and the irregular $\rm
SiO_2$ grains as well as of the resulting RBD agglomerates
thereof. The reference numbers refer to [A] manufacturer
information, micromod Partikeltechnologie GmbH, [B]
\citet{blusch04}, [C] \citet{popsch05}, [D] \citet{heietal99}, [E]
\citet{popetal00}, [F] \citet{bluwur00}, [G] manufacturer
information, Saint-Gobain Diamantwerkzeuge GmbH \& Co. KG, [H]
manufacturer information, Sigma-Aldrich Chemie GmbH, [I]
\citet{bluetal06}.} \tablehead{\colhead{Physical property} &
\colhead{Symbol} & \colhead{Value} & \colhead{Unit} &
\colhead{Reference}} \startdata
{\underline{Spherical $\rm SiO_2$ particles}}\\
Material && $\rm SiO_2$, non-porous && [A]\\
Morphology && spherical && [A]\\
Molecular arrangement && amorphous & & [A]\\
Density & $\rho_0$ & $2,000$ & $\rm kg~m^{-3}$ & [B]\\
Radius & $s_0$ & $0.76 \pm 0.03$ & $\rm \mu m$ & [C]\\
Mass & $m_0$ & $(3.7 \pm 0.4) \cdot 10^{-15}$ & kg &\\
Surface molecules && Si-OH && [A]\\
Surface energy && 0.014 & $\rm J~m^{-2}$ & [D]\\
Adhesion force & $F_{\rm stick}$ & $(67 \pm 11) \times 10^{-9}$ & N & [D]\\
Adhesion energy & $E_{\rm stick}$ & $(2.2 \pm 0.4) \times 10^{-15}$ & J & [E,F]\\
Rolling-friction force & $F_{\rm roll}$ & $(0.68 \pm 0.13) \times 10^{-9}$ & N & linear extrapolation from [D]\\
Rolling-friction energy & $E_{\rm roll}$ & $(8.1 \pm 1.9) \times 10^{-16}$ & J & $E_{\rm roll} = F_{\rm roll} \frac{\pi}{2} s_0$\\
Sticking threshold velocity & $v_{\rm stick}$ & 1.1 & $\rm m~s^{-1}$ & extrapolated from [E]\\
Rolling-threshold velocity & $v_{\rm roll}$ & $1.5 \pm 0.3$ & $\rm
m~s^{-1}$ & $v_{\rm roll} = \sqrt{(10 E_{\rm roll}) / m_0} $\\
\hline
{\underline{Agglomerates of spherical $\rm SiO_2$ particles}}\\
Volume filling factor & $\phi$ & $0.15 \pm 0.01$ & & [I]\\
Compressive strength & $C$ & 500 & $\rm N~m^{-2}$ &[I]\\
Tensile strength & $T$ & 1,100 & $\rm N~m^{-2}$ &[I]\\
\hline \hline
\\[5cm]
{\underline{Irregular diamond particles}}\\
Material && C, diamond && [G]\\
Morphology && irregular && \\
Density & $\rho_0$ & 3,520 & $\mathrm{kg\,m^{-3}}$ & \\
Size & $s_0$ & $0.75 \pm 0.25$ & $\rm \mu m$ & [E] \\
\hline
{\underline{Agglomerates of irregular diamond particles}}\\
Volume filling factor & $\phi$ & $0.11 \pm 0.02$ && [I]\\
Compressive strength & $C$ & 200 & $\rm N~m^{-2}$ &[I]\\
Tensile strength & $T$ & 200 & $\rm N~m^{-2}$ & [I]\\
\hline \hline
{\underline{Irregular $\rm SiO_2$ particles}}\\
Material && $\rm SiO_2$, non-porous && [H]\\
Morphology && irregular && \\
Density & $\rho_0$ & 2,600 & $\mathrm{kg\,m^{-3}}$ & [H] \\
Size & $s_0$ & $\sim$ 0.05-5 & $\rm \mu m$ & 50\% of typical particle diameter \\
\hline
{\underline{Agglomerates of irregular $\rm SiO_2$ particles}}\\
Volume filling factor & $\phi$ & $0.07 \pm 0.03$ &&[I]\\
Compressive strength & $C$ & 200 & $\rm N~m^{-2}$ &[I]\\
Tensile strength & $T$ & 300 & $\rm N~m^{-2}$ &[I]\\
\enddata
\end{deluxetable}

\end{document}